\documentclass[journal]{IEEEtran}
\usepackage{array}
\usepackage[caption=false,font=normalsize,labelfont=sf,textfont=sf]{subfig}
\usepackage{stfloats}
\usepackage{url}
\usepackage{verbatim}
\usepackage{pifont}
\usepackage{cite}
\usepackage{amsmath,amssymb,amsfonts}
\usepackage{graphicx}
\usepackage{textcomp}
\usepackage{algorithm}  
\usepackage{algpseudocode}  
\usepackage{booktabs}
\usepackage{multirow}
\graphicspath{{./imgs/}}

\def\BibTeX{{\rm B\kern-.05em{\sc i\kern-.025em b}\kern-.08em
    T\kern-.1667em\lower.7ex\hbox{E}\kern-.125emX}}
\begin{document}
\title{Hypernetwork-based Personalized Federated Learning for Multi-Institutional CT Imaging}
\author{Ziyuan Yang, Wenjun Xia, Zexin Lu, Yingyu Chen, Xiaoxiao Li, \IEEEmembership{Member, IEEE}, and Yi Zhang, \IEEEmembership{Senior Member, IEEE}
\thanks{Corresponding author: Yi Zhang}
\thanks{Ziyuan Yang, Wenjun Xia, Zexin Lu and Yingyu Chen are with the College of Computer Science, Sichuan University, Chengdu 610065, China (e-mail: cziyuanyang@gmail.com; xwj90620@gmail.com; zexinlu.scu@gmail.com; 2021323040005@stu.scu.edu.cn). }
\thanks{Xiaoxiao Li is with the Department of Electrical and Computer Engineering, The University of British Columbia, Vancouver, BC, V6T1Z4, Canada (e-mail: xiaoxiao.li@ece.ubc.ca)}
\thanks{Yi Zhang is with the School of Cyber Science and Engineering, Sichuan University, Chengdu 610065, China (e-mail: yzhang@scu.edu.cn)}

}

\maketitle

\begin{abstract}
Computed tomography (CT) is of great importance in clinical practice due to its powerful ability to provide patients' anatomical information without any invasive inspection, but its potential radiation risk is raising people’s concerns. Deep learning-based methods are considered promising in CT reconstruction, but these network models are usually trained with the measured data obtained from specific scanning protocol and need to centralizedly collect large amounts of data, which will lead to serious data domain shift, and privacy concerns. To relieve these problems, in this paper, we propose a hypernetwork-based federated learning method for personalized CT imaging, dubbed as HyperFed. The basic assumption of HyperFed is that the optimization problem for each institution can be divided into two parts: the local data adaption problem and the global CT imaging problem, which are implemented by an institution-specific hypernetwork and a global-sharing imaging network, respectively. The purpose of global-sharing imaging network is to learn stable and effective common features from different institutions. The institution-specific hypernetwork is carefully designed to obtain hyperparameters to condition the global-sharing imaging network for personalized local CT reconstruction. Experiments show that HyperFed achieves competitive performance in CT reconstruction compared with several other state-of-the-art methods. It is believed as a promising direction to improve CT imaging quality and achieve personalized demands of different institutions or scanners without privacy data sharing. The codes 
will be released at https://github.com/Zi-YuanYang/HyperFed.
\end{abstract}


\section{Introduction}
\label{sec:introduction}
\IEEEPARstart{C}{omputed} tomography (CT) is an important imaging modality in clinical diagnosis since it can noninvasively visualize anatomical information inside the patient’ body. However, with the popularization of CT, related concerns have been expressed about the potential radiation risk that may lead to genetic, cancerous and other diseases. To lower the radiation dose, switching the voltage/current of X-ray tube and reducing the scanning views are two commonly used strategies\cite{slovis2002alara}. However, both strategies will unavoidably degrade the imaging quality, which produce negative impacts on the subsequent image analysis and diagnosis \cite{bao2018few,chen2017low11,wu2021deep}.

Recently, attracted by the encouraging performance of deep learning (DL), researchers are enthusiastic about introducing DL to low-dose CT (LDCT) reconstruction and achieved impressive results \cite{wu2019computationally,ge2019stereo,wu2021low}. In spite of fruitful results obtained in recent years, current methods are all centralized training (CL)-based and need to collect huge amounts of training data from different institutions without considering privacy issue. Although these data are anonymously transferred, many works have proven that anonymization cannot effectively protect the patients' privacy \cite{narayanan2008robust,schwarz2019identification,gursoy2021functional}. Restricted to the privacy, legal and ethical concerns, since it is difficult to train a stable and effective model just based on a small number of local data, a privacy-preserving multi-institutional cooperative training method is highly demanded.


Federated learning (FL) is a recently proposed decentralized solution that aims to protect data privacy and confidentiality while the model can learn enough information from multiple data sources \cite{kaissis2020secure}.
The greatest difference between the training processes of CL-based and FL-based frameworks is the transferred contents. In CL-based methods, the patients' private data is transferred, but in FL-based methods, only gradients are transferred, and the privacy contained in the gradients is negligible compared with the contents transferred in CL-based methods.

One main challenge in FL lies in that the data from different sources are usually non-independent, identically distributed (non-iid), in which this problem is more severe in CT imaging than other analysis tasks. 
Due to the differences in both hardware and scanning protocol, the data collected from different scanners or institutions inevitably suffer from this problem. Unfortunately, the global model generated by FL methods can only capture some common statistical characteristics from different institutions and cannot provide specific features for different data sources. 


To relieve the non-iid problem, in this paper, we propose a novel hypernetwork-based personalized federated learning framework for multi-institutional CT imaging, dubbed as HyperFed. The basic assumption of this work is that the global optimization problem can be decomposed into two parts, the local data adaption and the global imaging. Since the CT imaging quality is highly related to the scanning protocol and geometry parameters\cite{xia2019spectral}, they can be considered as a personalization regulator to adapt data in CT imaging. The proposed HyperFed comprises two main modules: the institution-specific hypernetwork and the global-sharing imaging network. For each institution, corresponding scanning parameters are transformed as personalization features and fed into the institution-specific hypernetwork to generate hyperparameters to modulate the features from different data domains. Since the input dimension of the imaging network is much higher than the hypernetwork, it is hard for the imaging network to converge to a satisfactory solution only based on a small number of local data. One possible solution is to introduce the idea of FL and share the gradients globally to train a robust global model with the data from multiple sources. Institution-specific hypernetwork doesn’t participate in the global aggregation and  facilitate the personalization for different data domains, and the global-sharing imaging network is responsible for learning the global universal features from different institutions.

The main contributions of this paper can be summarized as follows:

(1) A novel hypernetwork-based personalized federated learning framework for multi-institutional CT imaging, dubbed as HyperFed, is proposed in this paper. To our best knowledge, this is the first attempt for personalized CT imaging based on FL.

(2) The proposed HyperFed is comprised of two main modules, the institution-specific hypernetwork and the global-sharing imaging network. The institution-specific hypernetwork is carefully designed for personalization to alleviate non-iid, and the primary purpose of global-sharing imaging network is to learn stable and effective common features from different institutions.

(3)	The proposed framework is flexible, which can be easily extended to different CT imaging tasks, such as post-processing and reconstruction.

\section{Related Works}
\subsection{CT Image Reconstruction}
CT imaging technology has witnessed remarkable innovations in the past decades and improved the diagnostic performance substantially. CT  reconstruction methods can be roughly divided into sinogram filtration, iterative reconstruction (IR) and post-processing.

Sinogram filtration methods work on either raw data or log-transformed data before image reconstruction, such as structural adaptive filtering \cite{balda2012ray}, bilateral filtering \cite{manduca2009projection} and penalized weighted least-squares \cite{wang2006penalized}. For most IR methods, the reconstruction problem is solved by optimizing an objective function formulated by combining the prior knowledge in sinogram and image domains \cite{donoho2006compressed}, such as the family of total variation (TV)\cite{sidky2008image,zhang2014few,sagheer2019denoising}, nonlocal means filter \cite{chen2009bayesian,zeng2015spectral,shim2021utility} and some other regularization terms \cite{xu2012low,chen2013improving,liu2021motion}. Post-processing-based methods, which do not need to access the raw data, are convenient to be deployed into current CT systems \cite{aharon2006k}. These methods achieve satisfactory performance if the inputs strictly follow their prior hypothesis, but they still suffer from high computational complexity and limited flexibility.

Recently, attracted by the impressive performance of deep learning (DL) in different fields, researchers are enthusiastic about introducing DL to LDCT reconstruction and achieved encouraging results. For example, Chen \textit{et al.} \cite{chen2017low} introduced the residual structure into the convolutional autoencoder and proposed a residual encoder-decoder convolutional neural network (RED-CNN) for low-dose denoising. You \textit{et al.} \cite{you2019ct} incorporated convolutional neural network (CNN), residual learning as well as generative adversarial network (GAN) techniques for CT super resolution. Yang \textit{et al.} \cite{yang2018low} trained a GAN with perceptual loss, which is defined as the distance between the reconstructed and the reference images in feature space and achieved satisfactory performance. Zhang \textit{et al.} \cite{zhang2021clear} extended GAN to reconstruct images from two domains by designing a comprehensive generator. Chen \textit{et al.} \cite{chen2018learn} unrolled the IR model optimized by gradient descent method into a network and learned the regularization terms and balancing parameters using training data. In \cite{chen2020airnet}, the modified proximal forward-backward splitting method was unrolled into residual reconstruction network corresponding to the updates of data fidelity and regularization terms. Xia \textit{et al.} \cite{xia2021ct} proposed a parameter-dependent framework (PDF) to introduce the scanning parameters into CT image denoising/reconstruction and achieved remarkable improvement. In spite of fruitful results obtained in recent years, current models are all CL-based, which is at a risk of privacy leakage during the process of data collection and severely restricts the promotion of DL-based methods in real clinical scenarios.

\subsection{Federated Learning}
Federated learning is a recently proposed decentralized solution that aims to protect data privacy and confidentiality while the model can learn enough information from multiple data sources \cite{zhang2021subgraph}. Typically, McMahan \textit{et al.} \cite{mcmahan2017communication} proposed FedAvg, which learns the global model by averaging local models of different parties. FedProx \cite{li2020federated} is considered as improved FedAvg, which constrains the local models close to the global model. Similarly, Li \textit{et al.} \cite{li2021model} proposed the model-contrastive (MOON) method to constrain the local models close to the global model by minimizing the contrastive distances between them. However, FL-based methods suffer from non-iid problem. To relieve this problem, several works were witnessed in personalized FL. Li \textit{et al.} \cite{li2021fedbn} proposed FedBN, which alleviates the feature shift using local batch normalization to achieve personalization in local institutions. In \cite{arivazhagan2019federated}, the network is composed of base and personalization layers. The base layers are trained following \cite{mcmahan2017communication} and the personalization layers are trained with local data and not delivered to the server for aggregation. Hanzely and Richtárik \cite{hanzely2020federated} added the regularization term to calculate the distance between local models and the global model to control the optimization degree.

Benefited from the privacy-preserving characteristic, FL was introduced in different medical tasks in recent years \cite{kaissis2021end,wu2021federated,park2021federated}. We also notice that there are some contemporary works about medical imaging. For example, Guo \textit{et al.} \cite{guo2021multi} proposed an intermediate latent feature alignment method for magnetic resonance image (MRI) reconstruction. Feng \textit{et al.} \cite{feng2021specificity} proposed a personalized model for MRI denoising, which shares the encoder rather than the whole model. However, CT reconstruction is heavily impacted by the scanning protocol and geometry parameters, which is quite different from MRI reconstruction and leads to a serious non-iid problem in federated learning for CT reconstruction. To relieve this problem, in this paper, we propose a personalized CT reconstruction framework. To our best knowledge, it is the first attempt in this field. Furthermore, the proposed model is flexible which can be easily extended to different CT imaging tasks, such as post-processing and reconstruction.

\section{Methodology}
\subsection{Models of CT Reconstruction and Personalized Federated Learning}
In general, CT reconstruction can be formulated as \cite{wu2021drone}:

\begin{equation}
\underset{x}{\operatorname{argmin}} \frac{1}{2}\|A x-y\|_{2}^{2}+R(x),\label{eq}
\end{equation}
where $\|\cdot\|_{2}^{2}$ is the $L_{2}$ norm and $A$ represents the system matrix. $x$ and $y$ denote the image be reconstructed and the measured data, respectively. $R(\cdot)$ denotes the regularization term, which is usually formulated with prior knowledge.

Besides, LDCT image denoising can be formulated as \cite{bai2021probabilistic}:

\begin{equation}\underset{f}{\operatorname{argmin}}\|f(x_l)-x_n\|_{2}^{2},\end{equation}
where $x_l$ is the low-dose image and $x_n$ is the corresponding normal-dose image. $f$ is the objective function to push $x_l$ to $x_n$ as close as possible.

For personalized FL, assuming that there are $K$ institutions, the learning process can be formulated as \cite{mansour2020three}:

\begin{equation}
\underset{\delta_{1}, \ldots, \delta_{K}}{\operatorname{argmin}}\left(\sum_{k=1}^{K} p_{k} F_{k}\left(\delta_{k}, \mathcal{D}^{k}\right)\right),
\end{equation}
where $F_k$ represents the optimization model in the $k$-th institution, $p_k$ stands for the weight of the $k$-th institution in global optimization, and $D^k$ denotes the dataset in the $k$-th institution. $\delta_k$ denotes the parameters of $F_k$. The purpose of personalized FL is to search the optimal local models for different institutions without data sharing.

\begin{figure}[!t]
\centerline{\includegraphics[width=\columnwidth]{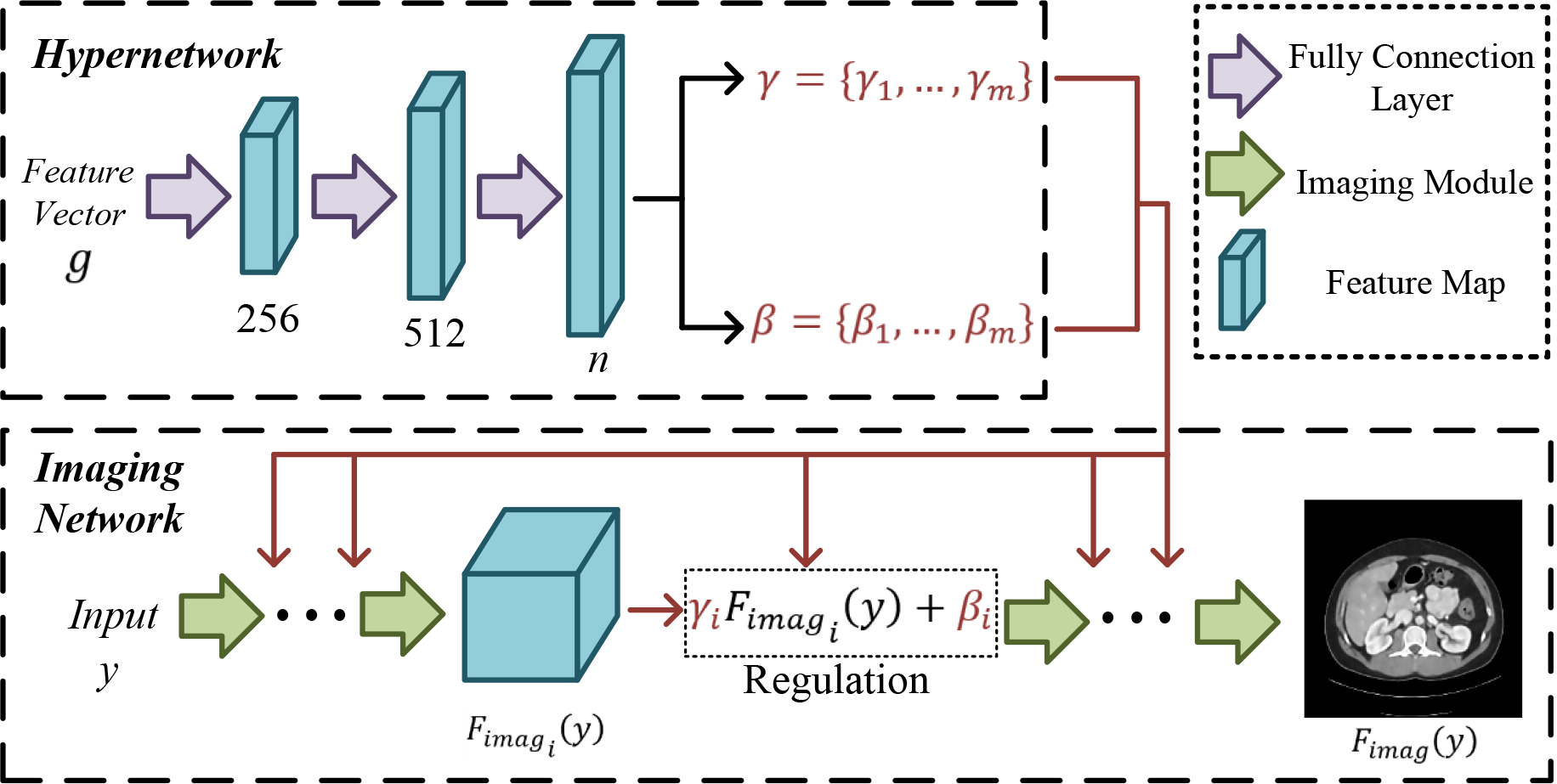}}
\caption{The architecture of the proposed HyperFed.}
\label{fig3}
\end{figure}

\subsection{The Architecture of HyperFed}

Since scanning protocol and geometry parameters heavily impact the CT reconstruction, it is reasonable to consider that these parameters contain information to guide the imaging network to predict the normal-dose CT. This idea motivates us to take full advantage of these parameters to alleviate non-iid problem and improve imaging performance. Specifically, the architecture of HyperFed is composed of an institution-specific hypernetwork and a global-sharing imaging network. The learning process of current imaging networks can be considered as $F_{imag}(x,y,w)$. The inputs are fed into the model $F_{imag}$ parameterized with $w$, and no constraint is applied. It is a common trick to reduce the complexity when all the inputs strictly follow a uniform distribution. However, it is hard to meet this assumption in real situations and non-iid problem usually appears. To deal with this problem, the institution-specific hypernetwork is implemented as a regulator to modulate the learning process. Then, the learning process of HyperFed is reformulated as $F_{imag}(x,y,w,\theta)$, where $\theta$ is the hyperparameters learned from the hypernetwork $F_{hyper}$ parameterized with $\xi$. Specifically, the feature vector $g$, which contains the scanning and geometry parameters, is fed into $F_{hyper}$ to generate the regularization factor set $\gamma$ and the bias set $\beta$, which are used to modulate the features of $F_{imag}$. This process is formulated as:

\begin{equation}
\gamma, \beta=F_{hyper}(g, \xi).
\end{equation}
For simplicity, in this paper, $g$ includes the number of detector bins, pixel length (mm), detector bin length (mm), the distance between the source and rotation center (mm), the distance between the detector and rotation center (mm) as well as the photon number of incident X-rays. Normalization and logarithmization are performed for some elements in $g$ with large magnitudes, such as the numbers of view and detector bins and the photon number of incident X-rays. The normalization is defined as:

\begin{equation}
    g_{i}=\frac{g_{i}-\min \left(g_{i}\right)}{\max \left(g_{i}\right)-\min \left(g_{i}\right)},
\end{equation}
where $i$ is the element index in $g$. $\max(\cdot)$ and $\min(\cdot)$ represent the maximum and minimum functions, respectively.

\begin{figure}[!t]
\centerline{\includegraphics[width=\columnwidth]{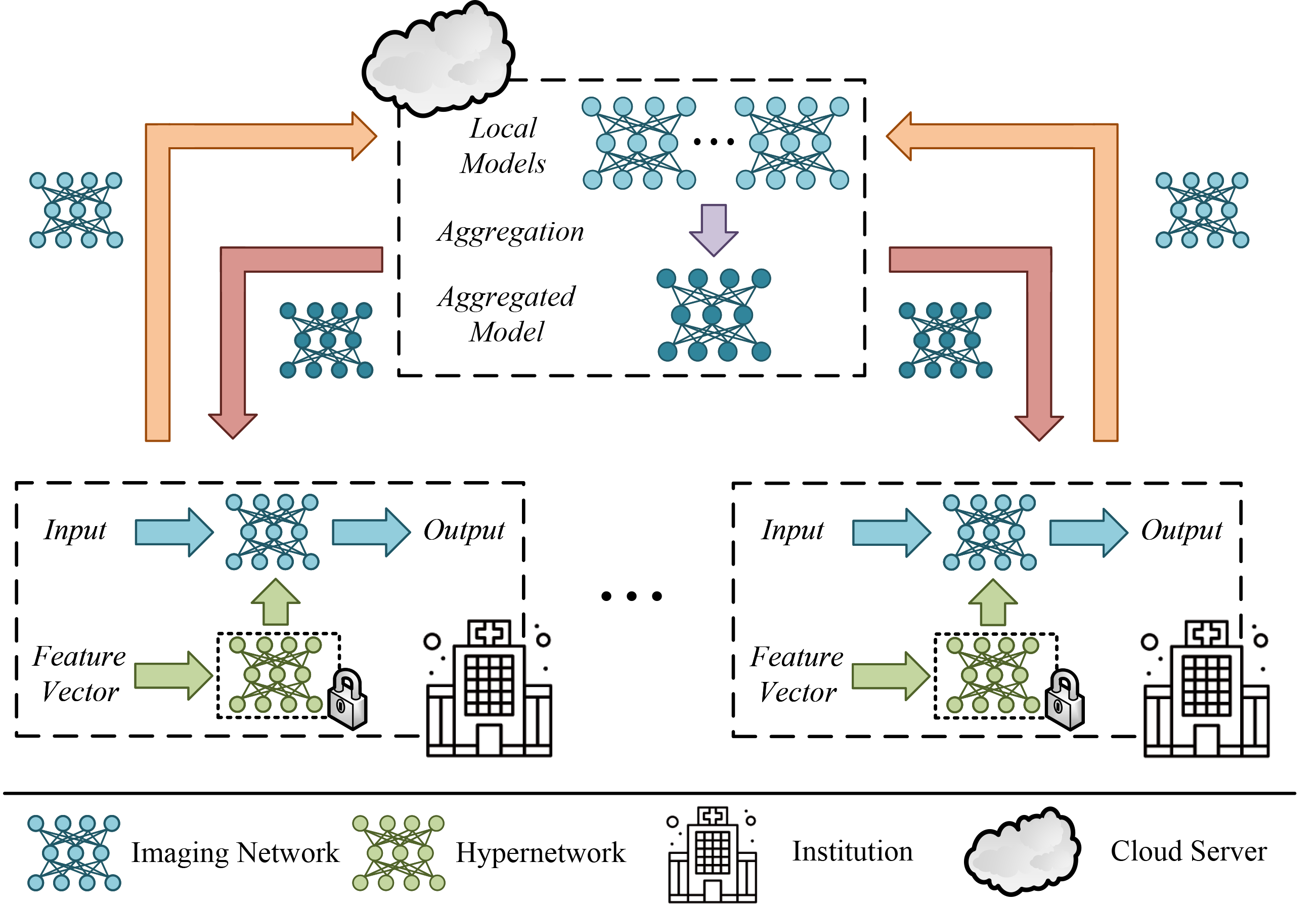}}
\caption{The training process of the proposed HyperFed.}
\label{fig4}
\end{figure}

Fig. \ref{fig3} shows the architecture of our proposed HyperFed. In Fig. \ref{fig3}, the output dimension of the hypernetwork is set according to the dimension of the feature maps from the imaging network. The outputs of the hypernetwork are utilized to modulate the feature maps of the CT imaging network. Specifically, the hypernetwork is a simple two-layer fully-connected network, which is used to generate $\gamma$ and $\beta$. The modulation function can be formulated as follows:

\begin{equation}
F_{imag}^{*}(y)=\gamma F_{imag}(y)+\beta,
\end{equation}

The modulation operation is applied on the feature maps from different modules in the imaging network and (6) is reformulated as:

\begin{equation}
    F_{{imag}_{i}}^{*}(y)=\gamma_{i} F_{{imag}_{i}}(y)+\beta_{i},
\end{equation}
where $F_{{imag}_{i}}(y)$ and $F_{{imag}_{i}}^{*}(y)$ represent the feature map from the $i$-th module and its corresponding modulated feature map, respectively. $\gamma_i$ and $\beta_i$ represent the regularization factor and bias for the $i$-th module, respectively.

As we mentioned above, the imaging network in the proposed HyperFed is flexible for different tasks. Imaging units in the proposed architecture vary for different imaging methods. For example, the imaging unit can be convolution layers for the post-processing method, such as RED-CNN \cite{chen2017low}. On the other side, for the unrolled iteration methods, such as LEARN \cite{chen2018learn}, it denotes an unrolled iteration module.

\begin{table*}[!t]
    \centering
    \setlength{\tabcolsep}{1.5mm}{
    \caption{The Geometry Parameters and Dose Levels in Different Institutions for Different Tasks (Post-processing $\backslash$ Reconstruction).}
    \label{tb1}
    \begin{tabular}{cccccc} 
    \toprule
                        & Institution \#1 & Institution \#2 & Institution \#3 & Institution \#4 & Institution \#5\\
      \midrule
Number of views                                          & 512 $\backslash$ 1024            & 512 $\backslash$ 88             & 384 $\backslash$ 1024             & 400 $\backslash$ 128             & 384 $\backslash$ 108             \\
Number of detector bins                                  & 368 $\backslash$ 512            & 315 $\backslash$ 768             & 330 $\backslash$ 768             & 350 $\backslash$ 512             & 350 $\backslash$ 512             \\
Pixel length (mm)                                        & 1.33 $\backslash$ 0.66            & 1.40 $\backslash$ 0.78            & 1.39 $\backslash$ 1.00            & 1.20  $\backslash$ 1.20           & 1.40  $\backslash$ 0.50            \\
Detector bin length (mm)                                 & 2.57 $\backslash$ 0.72            & 3.00 $\backslash$ 0.58            & 2.60 $\backslash$ 0.62            & 2.20 $\backslash$ 1.40            & 2.50 $\backslash$ 0.40            \\
Distance between the source and rotation center (mm)     & 595 $\backslash$ 250             & 450 $\backslash$ 350             & 400 $\backslash$ 500             & 400 $\backslash$ 500             & 500 $\backslash$ 400             \\
Distance between the detector and rotation center   (mm) & 491 $\backslash$ 250             & 350 $\backslash$ 300             & 300 $\backslash$ 400             & 350 $\backslash$ 500             & 300 $\backslash$ 200             \\
Intensity of X-rays                                      & 0.5e5 $\backslash$ 1e5           & 0.6875e5 $\backslash$ 1e6        & 0.875e5 $\backslash$ 5e4         & 1.0625e5 $\backslash$ 2.5e5        & 1.25e5 $\backslash$ 5e5          \\ 
      \bottomrule
      \end{tabular}}
\end{table*}

  \begin{algorithm}[htb]  
  \caption{Main steps of HyperFed.}  
  \label{alg:Framwork}  
  \begin{algorithmic}[1]  
\Require $\mathcal{D} \triangleq U_{k \in K} \mathcal{D}^{k}$, data from $K$ institutions; $T$, the number of communication rounds; $E$, the number of local epochs.
    \Ensure The optimized parameters $w$ of the imaging network, and $\xi_k$ of the the hypernetwork in the $k$-th institution.
    \State \textbf{Server executes:}
    \State Randomly initialize $w^{init}$ and $\xi^{init}$ on the server and deliver them to each institution.
     \For{$t=1,2,...,T$}
        \For{$k=1,2,...,K$ \textbf{in parallel}} 
        \State send $w^t$ to $k$-th institution
        \State $w_k^t \gets $ \textbf{Institution Local Training}($k$,$w^t$)
        \EndFor\\
        \hspace{1.5em}$w^{t+1} \gets \sum_{k=1}^{K} \frac{\left|\mathcal{D}^{k}\right|}{|\mathcal{D}|} w_{k}^{t}$
     \EndFor
    \State \textbf{Institution Local Training:}
    \State $w_k^t\gets w^t$
    \For{$e=1,2,...,E$}
        \For{$(y,g,\hat{x})$ in $D^k$}
        \State $\gamma, \beta \gets F_{hyper}^{k}(g)$
        \State $l\gets MSELoss\left(\gamma F_{i m a g}^{k}(y)+\beta, \hat{x}\right)$
        \State $w_{k}^{(t, e+1)}, \xi_{k}^{(t, e+1)}\gets Adam\left(w_{k}^{(t, e)}, \xi_{k}^{(t, e)}\right)$
        \EndFor
    \EndFor \\
    \Return $w_k^{(t,E)}$ to the server
  \end{algorithmic}  
\end{algorithm}

\subsection{Implementation of HyperFed}

In this paper, we propose a hypernetwork-based personalized  FL framework to relieve non-iid problem in CT image reconstruction. Similar to other federated learning methods, such as FedAvg and FedProx, HyperFed updates the hypernetwork and imaging network locally and only averages the imaging network in the server. The training process is depicted in Fig. \ref{fig4}. Similar to FedBN, the data normalization method is performed at local in HyperFed. We notice that the proposed HyperFed is expected to predict the reconstruction results for different scanners from different institution, whose data follows different distributions \cite{lim2017enhanced} and is against the assumption of FedBN that batch normalization is helpful if the output obeys a specific distribution. This mismatching will compromise the model performance. To circumvent this obstacle, the hypernetwork is proposed to modulate the feature maps of the imaging network, which can be approximately considered as self-normalization.

As discussed above, the cooperation among different institutions is hard to implement due to the privacy and security concerns in the real situation. HyperFed is proposed to allow different institutions to cooperate in training an imaging network in a privacy-preserving way. In HyperFed, local data can only be accessed by its own institution, and only the local gradients are aggregated in the server, which can effectively protect patients’ privacy. In the proposed HyperFed, the model is initialized in the server and delivered to different institutions at first. For each institution, the local model is trained with its own data by minimizing the following loss:

\begin{equation}
    \mathcal{L}^{k}=E_{(y, g, \hat{x}) \sim \mathrm{p}\left(\mathcal{D}^{k}\right)} \frac{1}{2}\left\|F^{k}(\delta_{k},y, g)-x_n\right\|_{2}^{2},
\end{equation}
where $F^k(\cdot)$ denotes the local network (including both hypernetwork and imaging network) at the $k$-th institution paramterized by $\delta_k$.


Assuming $w_k$ and $\xi_{k}$ represent the parameters of $F_{image}$ and $F_{hyper}$ at the $k$-th institution, respectively, the parameter optimization can be formulated as:
\begin{equation}
    w_{k}^{p+1}, \xi_{k}^{p+1}=\left(w_{k}^{p}, \xi_{k}^{p}\right)-\lambda \nabla_{\left(w_{k}^{p}, \xi_{k}^{p}\right)} \mathcal{L}^{k},
\end{equation}
where $w_k^p$ and $\xi_k^p$ represents the $p$-th training epoch of $w_k$ and $\xi_k$,  and $\lambda$ stands for the learning rate.

After every several training epochs, the gradient of the imaging network will be delivered to the server for aggregation, and the hypernetwork is preserved at local for modulation. The main steps of our proposed HyperFed are listed in Algorithm 1. Adam \cite{kingma2014adam} is used to optimize the network, and mean squard error (MSE) is adopted as the loss function.

\section{Experiments}

\subsection{Experimental Setting}

\begin{figure}[!t]
	\centering
	\begin{minipage}[t]{0.18\columnwidth}
	\centering
	\includegraphics[width=\columnwidth]{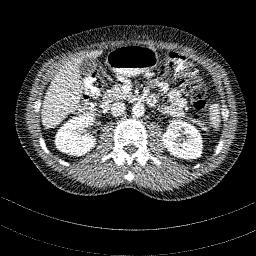}
	\centerline{(a)}
	\end{minipage}
	\begin{minipage}[t]{0.18\columnwidth}
	\includegraphics[width=\columnwidth]{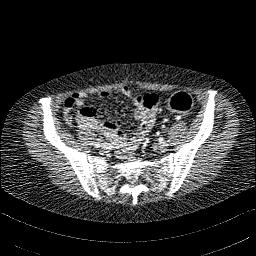}
	\centerline{(b)}
	\end{minipage}
	\begin{minipage}[t]{0.18\columnwidth}
	\includegraphics[width=\columnwidth]{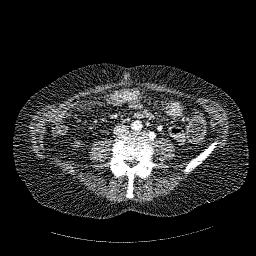}
	\centerline{(c)}
	\end{minipage}
	\begin{minipage}[t]{0.18\columnwidth}
	\includegraphics[width=\columnwidth]{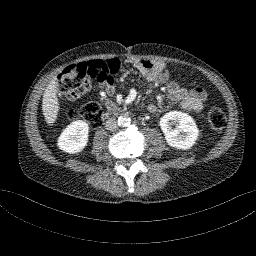}
	\centerline{(d)}
	\end{minipage}
	\begin{minipage}[t]{0.18\columnwidth}
	\includegraphics[width=\columnwidth]{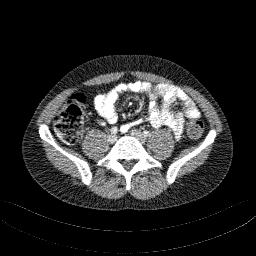}
	\centerline{(e)}
	\end{minipage}\\
	\begin{minipage}[t]{0.18\columnwidth}
	\centering
	\includegraphics[width=\columnwidth]{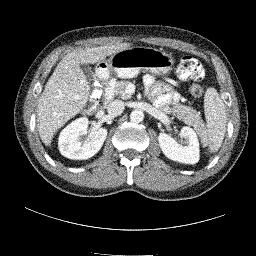}
	\centerline{(f)}
	\end{minipage}
	\begin{minipage}[t]{0.18\columnwidth}
	\includegraphics[width=\columnwidth]{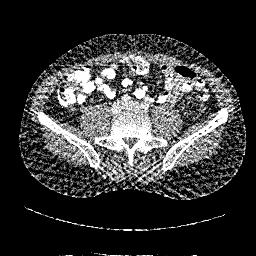}
	\centerline{(g)}
	\end{minipage}
	\begin{minipage}[t]{0.18\columnwidth}
	\includegraphics[width=\columnwidth]{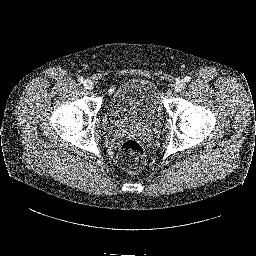}
	\centerline{(h)}
	\end{minipage}
	\begin{minipage}[t]{0.18\columnwidth}
	\includegraphics[width=\columnwidth]{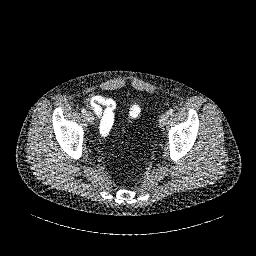}
	\centerline{(i)}
	\end{minipage}
	\begin{minipage}[t]{0.18\columnwidth}
	\includegraphics[width=\columnwidth]{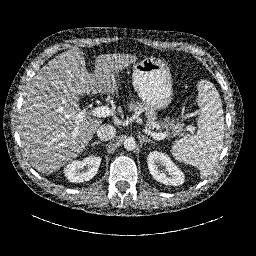}
	\centerline{(j)}
	\end{minipage}
\caption{The simulated samples. (a)-(e) represent the simulated samples in different institutions for the post-processing task; (f)-(j) represent the simulated samples in different institutions for the reconstruction task. The display window is [-160, 240] HU.}
\label{fig5}
\end{figure}

\begin{figure*}[!t]
\centerline{\includegraphics[width=0.88\textwidth]{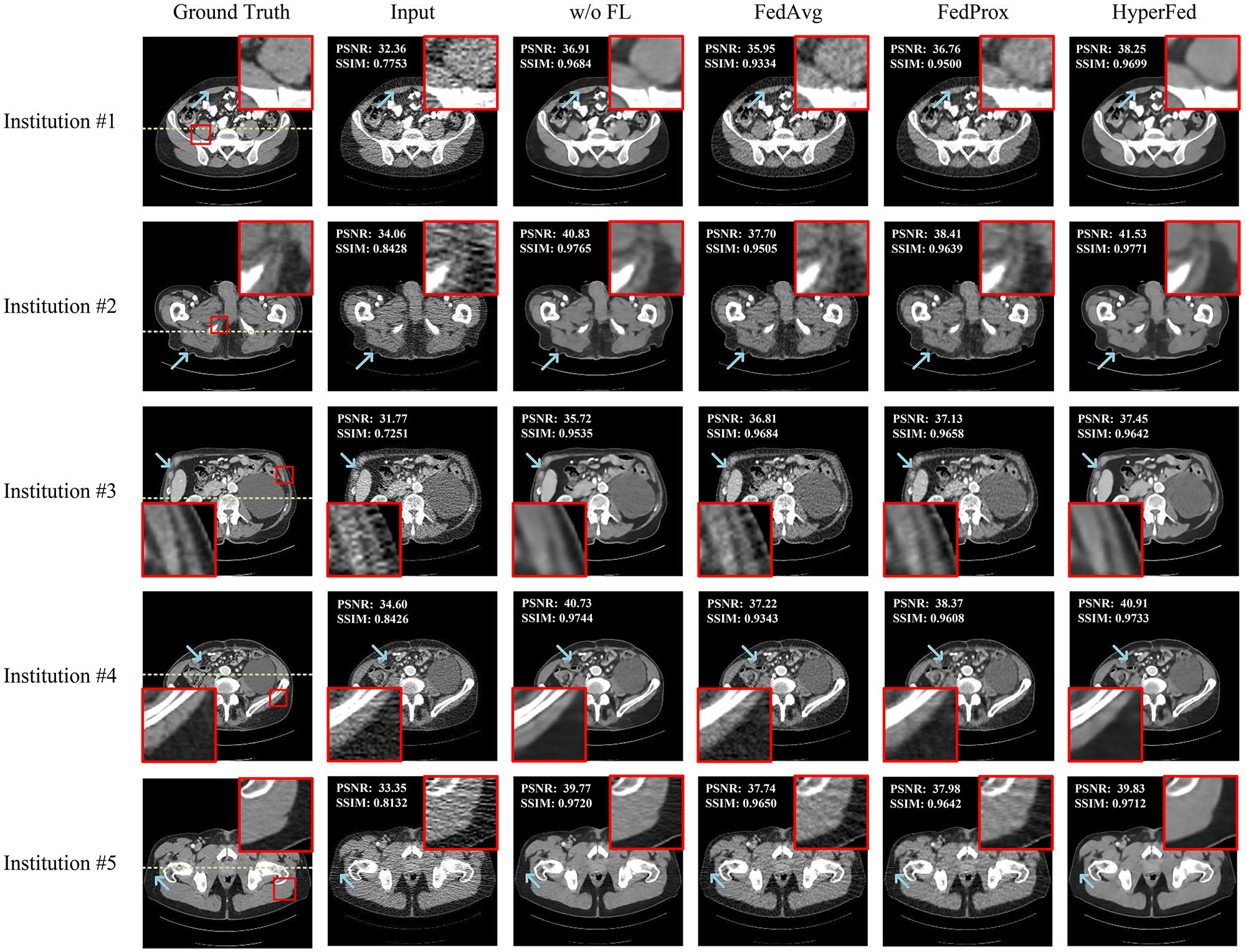}}
\caption{The results of w/o FL, FedAvg, FedProx and HyperFed under different geometries and dose levels for the post-processing task. The display window is [-160, 240] HU.}
\label{fig6}
\end{figure*}

\begin{figure}[!t]
	\centerline{\includegraphics[width=\columnwidth]{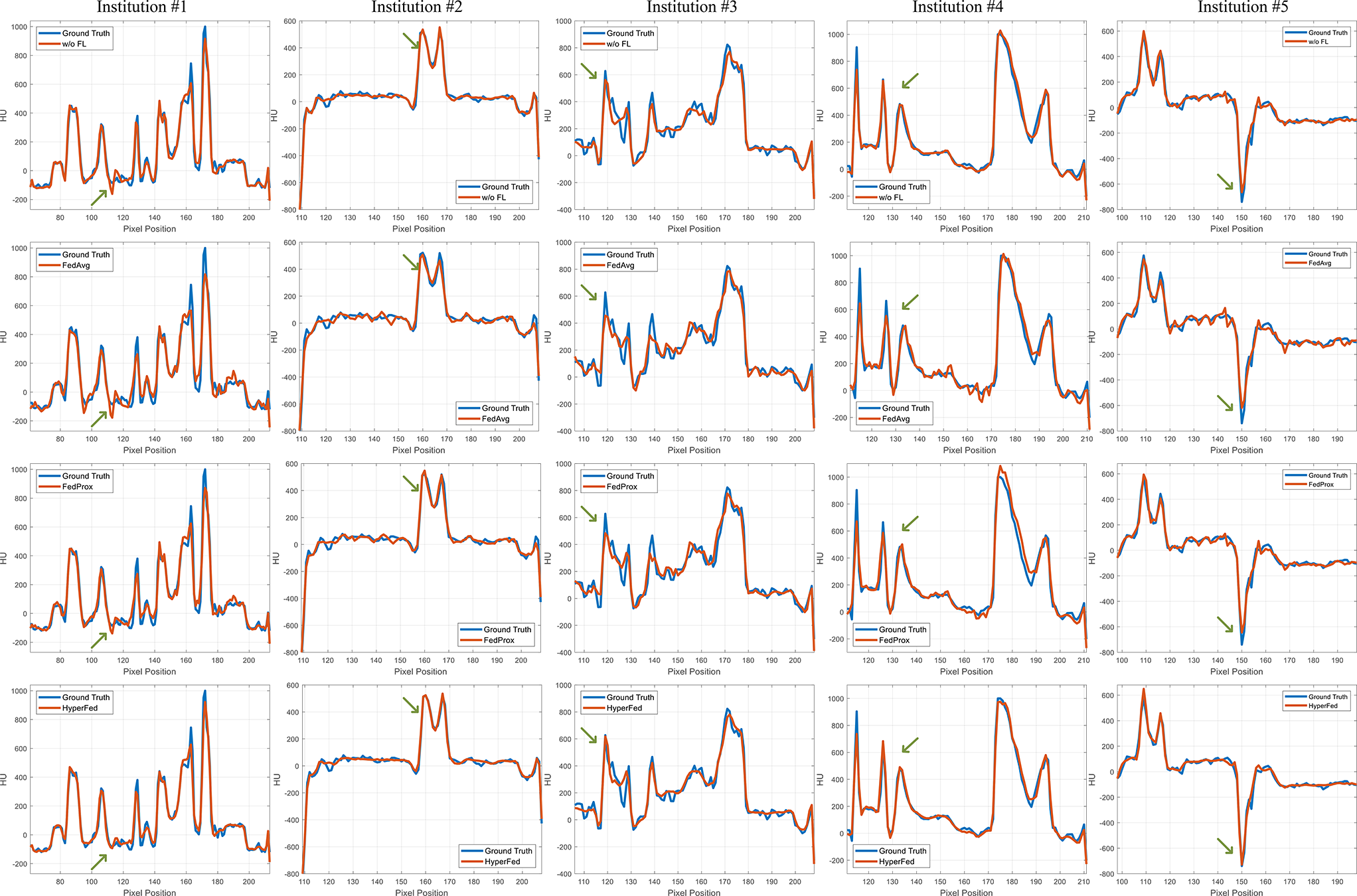}}
	\caption{Horizontal profiles of the results of w/o FL, FedAvg, FedProx and HyperFed.}
\label{fig7}
\end{figure}

\begin{figure}[!t]
	\centering
	\begin{minipage}[t]{0.32\columnwidth}
	\centering
	\includegraphics[width=\columnwidth]{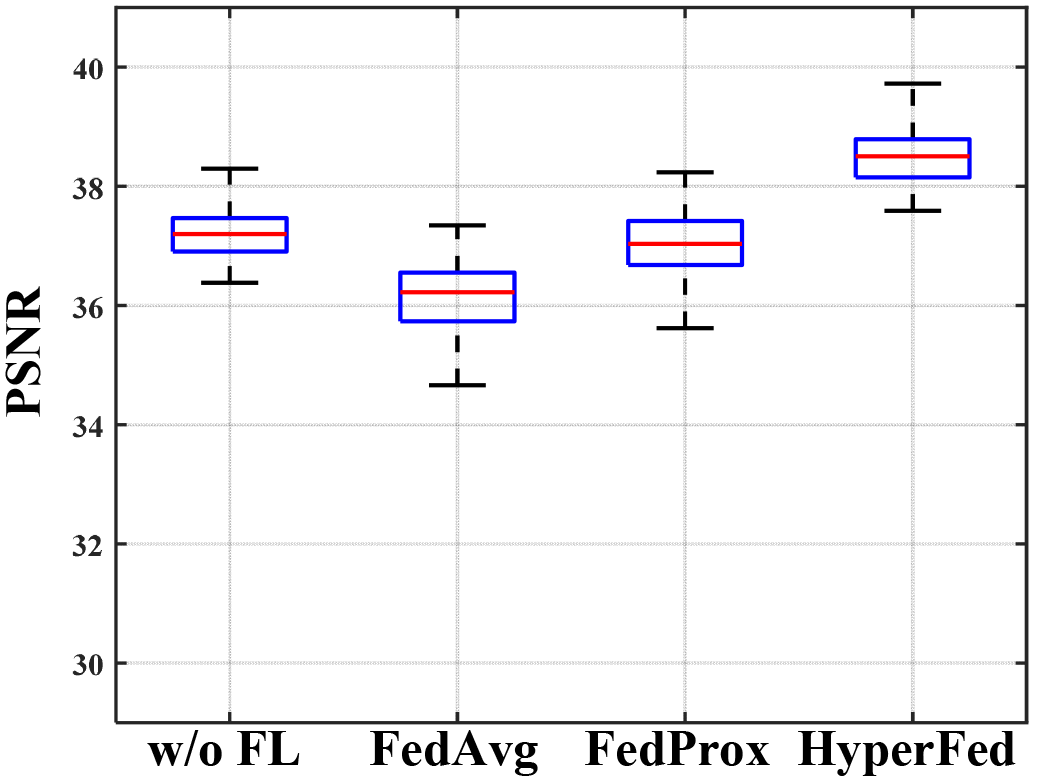}
	\centerline{(a)}
	\end{minipage}
	\begin{minipage}[t]{0.32\columnwidth}
	\includegraphics[width=\columnwidth]{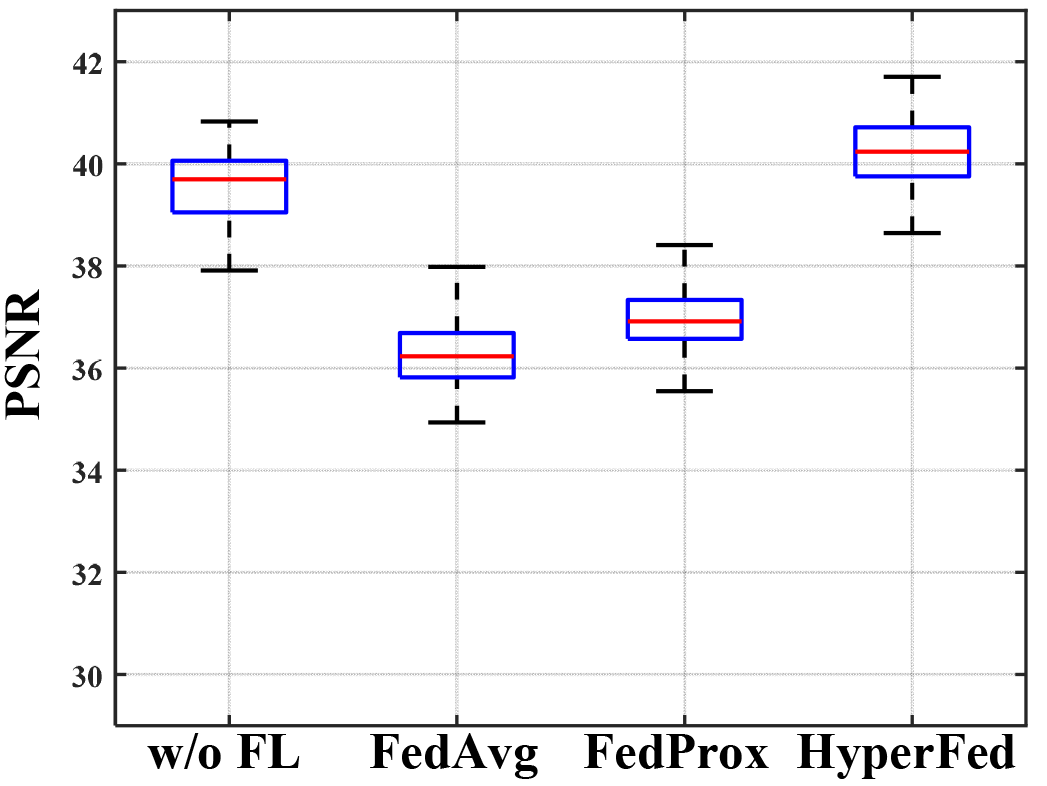}
	\centerline{(b)}
	\end{minipage}
	\begin{minipage}[t]{0.32\columnwidth}
	\includegraphics[width=\columnwidth]{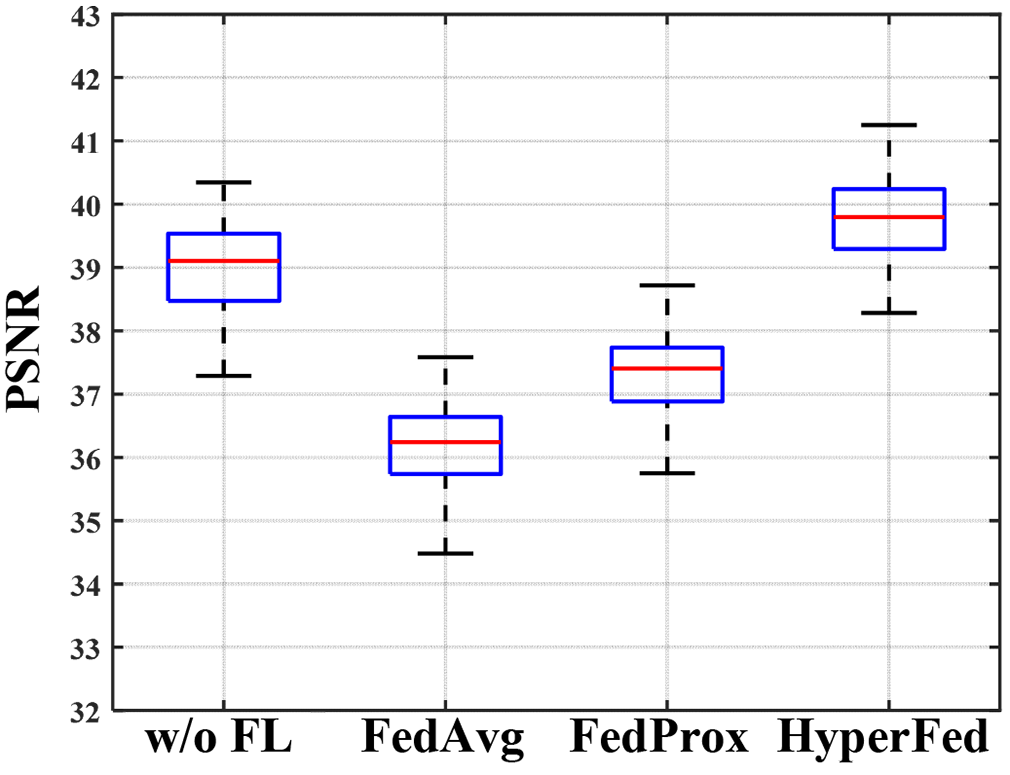}
	\centerline{(c)}
	\end{minipage}
\\
	\begin{minipage}[t]{0.32\columnwidth}
	\centering
	\includegraphics[width=\columnwidth]{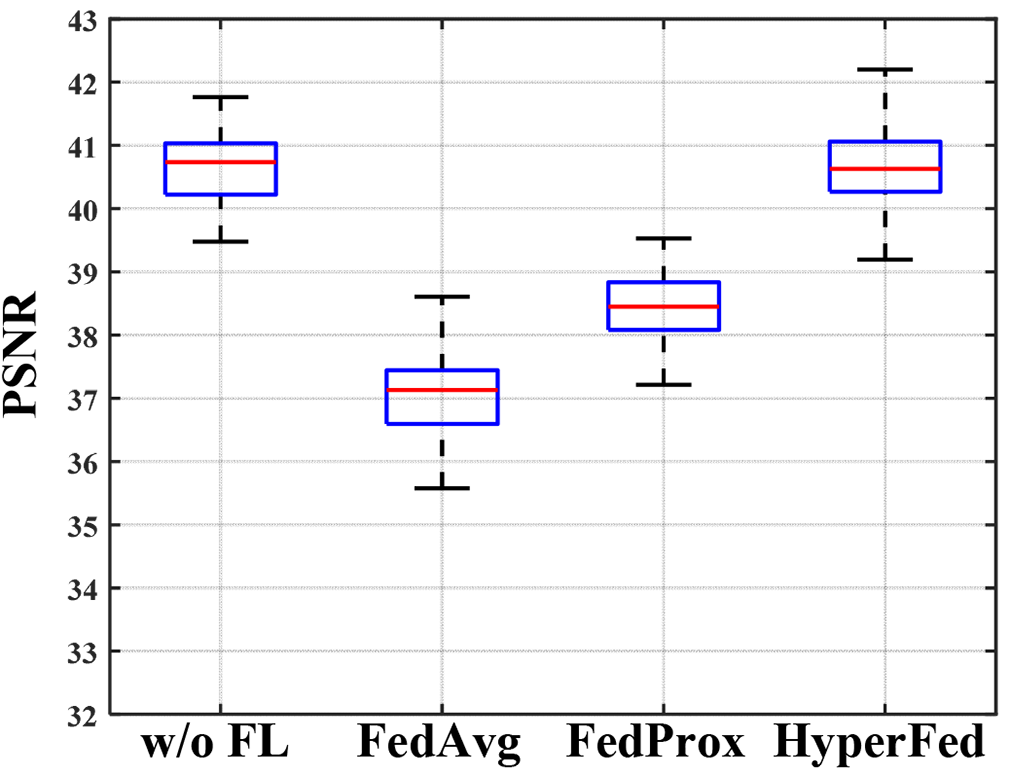}
	\centerline{(d)}
	\end{minipage}
	\begin{minipage}[t]{0.32\columnwidth}
	\includegraphics[width=\columnwidth]{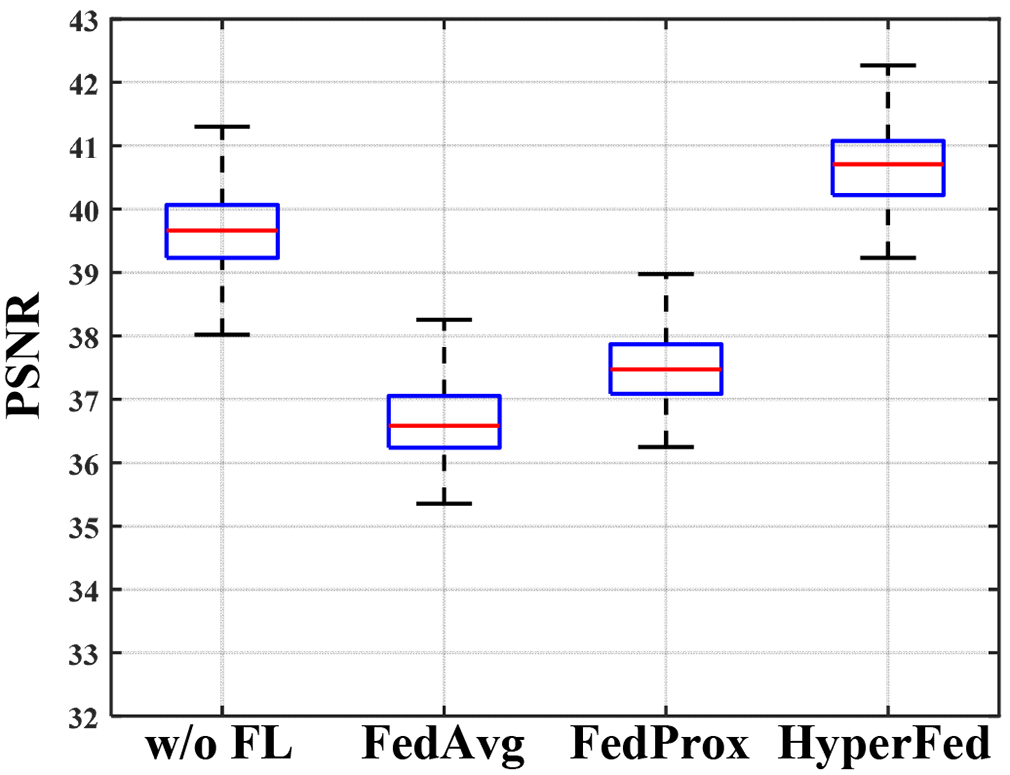}
	\centerline{(e)}
	\end{minipage}
	\begin{minipage}[t]{0.32\columnwidth}
	\includegraphics[width=\columnwidth]{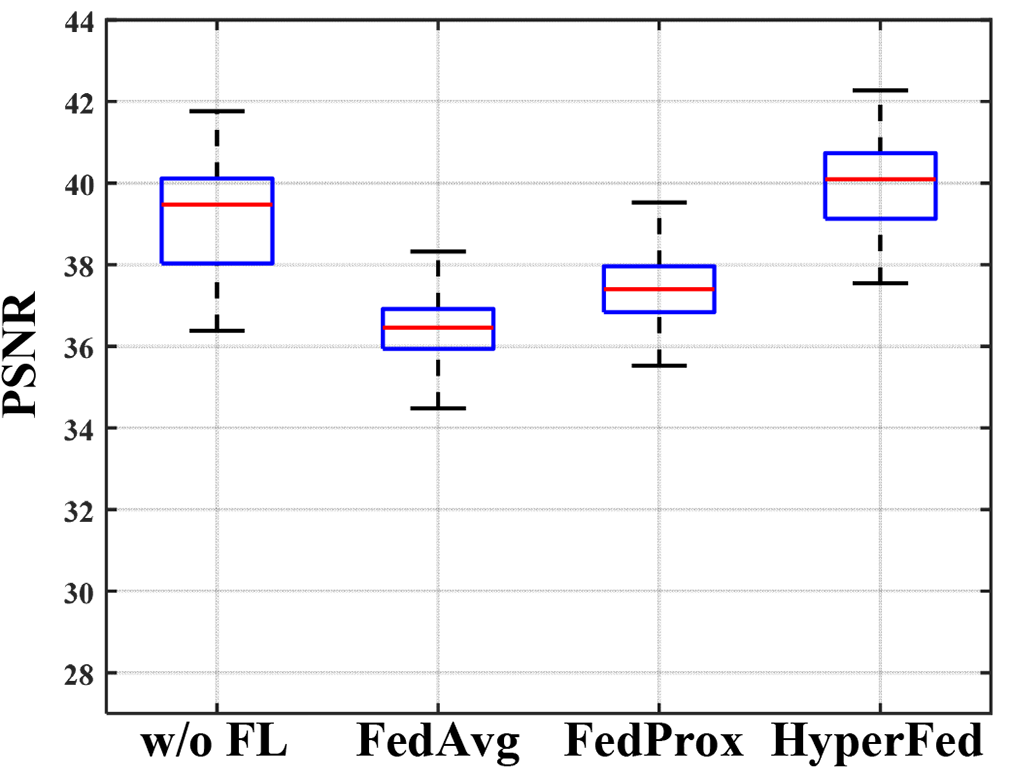}
	\centerline{(f)}
	\end{minipage}
\caption{The boxplots of PSNR based on w/o FL, FedAvg, FedProx and HyperFed for the post-processing task. (a)-(g) represent the results of institution \#1 to \#5 respectively and (f) represents the average result of all institutions.}
\label{fig8}
\end{figure}

\begin{table*}[]
\centering
\caption{The Quantitative Results of PSNR and SSIM for the Past-processing Task.}
\label{tb3}
\begin{tabular}{@{}lllllllll@{}}
\toprule
\centering
\multirow{2}{*}{} & \multicolumn{2}{c}{w/o   FL} & \multicolumn{2}{c}{FedAvg} & \multicolumn{2}{c}{FedProx} & \multicolumn{2}{c}{HyperFed} \\
                  & PSNR         & SSIM          & PSNR        & SSIM         & PSNR         & SSIM         & PSNR         & SSIM          \\ \midrule
Insitution \#1    & 36.97        & 0.9506        & 35.94       & 0.9132       & 36.78        & 0.9317       & \textbf{38.16}        & \textbf{0.9518}        \\
Insitution \#2    & 39.21        & 0.9544        & 36.02       & 0.9211       & 36.65        & 0.9342       & \textbf{39.84}        & \textbf{0.9555}        \\
Insitution \#3    & 38.68        & 0.9512        & 35.96       & 0.9039       & 37.06        & 0.9317       & \textbf{39.45}        & \textbf{0.9530}        \\
Insitution \#4    & 40.25        & \textbf{0.9614}        & 36.77       & 0.9174       & 38.16        & 0.9458       & \textbf{40.26}        & 0.9581        \\
Insitution \#5    & 39.31        & 0.9551        & 36.38       & 0.9224       & 37.20        & 0.9393       & \textbf{40.23}        & \textbf{0.9566}        \\
Overall           & 38.88        & 0.9545        & 36.21       & 0.9156       & 37.17        & 0.9365       & \textbf{39.59}        & \textbf{0.9550}        \\ \bottomrule
\end{tabular}
\end{table*}

The “2016 NIH-AAPM-Mayo Clinic Low-Dose CT Grand Challenge” dataset \cite{mccollough2016tu}, which contains 5936 full-dose CT images from 10 patients, is used to evaluate the proposed method. 400 and 100 images are randomly selected as the training and testing datasets, respectively. These images are randomly divided into five groups on average to simulate the images with different scanning parameters. In order to demonstrate the generalization of the proposed method, HyperFed is validated on different CT imaging tasks, including post-processing and reconstruction. Considering different application scenarios, we simulate the datasets according to our previous work \cite{xia2021ct}. For post-processing and reconstruction tasks, two different sets of geometric parameters and dose levels are used for simulation, which are listed in Tab. \ref{tb1}, respectively. For the post-processing task, HyperFed is validated on the data with different dose levels, and for the reconstruction task, HyperFed is tested on the hybrid data, including both sparse-view and low-dose cases. Fig. \ref{fig5} demonstrates several simulated examples. It is seen that different parts of the human torso are included and the noise and artifacts are different in each case. Peak signal-to-noise ratio (PSNR) and structural similarity index measure (SSIM) are adopted as the quantitative metrics.

For simplicity, we assume that each institution only has one type of data, and the data transmission is strictly prohibited. RED-CNN \cite{chen2017low} and LEARN \cite{chen2018learn}, which are one of the most representative methods in post-processing and unrolled reconstruction networks are adopted, respectively. The learning rate is set to $1\times10^{-4}$, the number of local training epochs is set to 3, and the number of communication rounds are set to 600 and 200 for RED-CNN and LEARN, respectively. We compare the proposed HyperFed with FedAvg \cite{mcmahan2017communication}, FedProx \cite{li2020federated} and the original imaging models without federated learning, which is dubbed as w/o FL. For FedProx, the penalty constant hyperparameter is set as $1\times10^{-4}$. All codes are implemented in PyTorch and the experiments are performed on a NVIDIA GTX 3090 GPU.

\subsection{Experiments for the Post-Processing Task}

\begin{figure*}[!t]
\centerline{\includegraphics[width=0.88\textwidth]{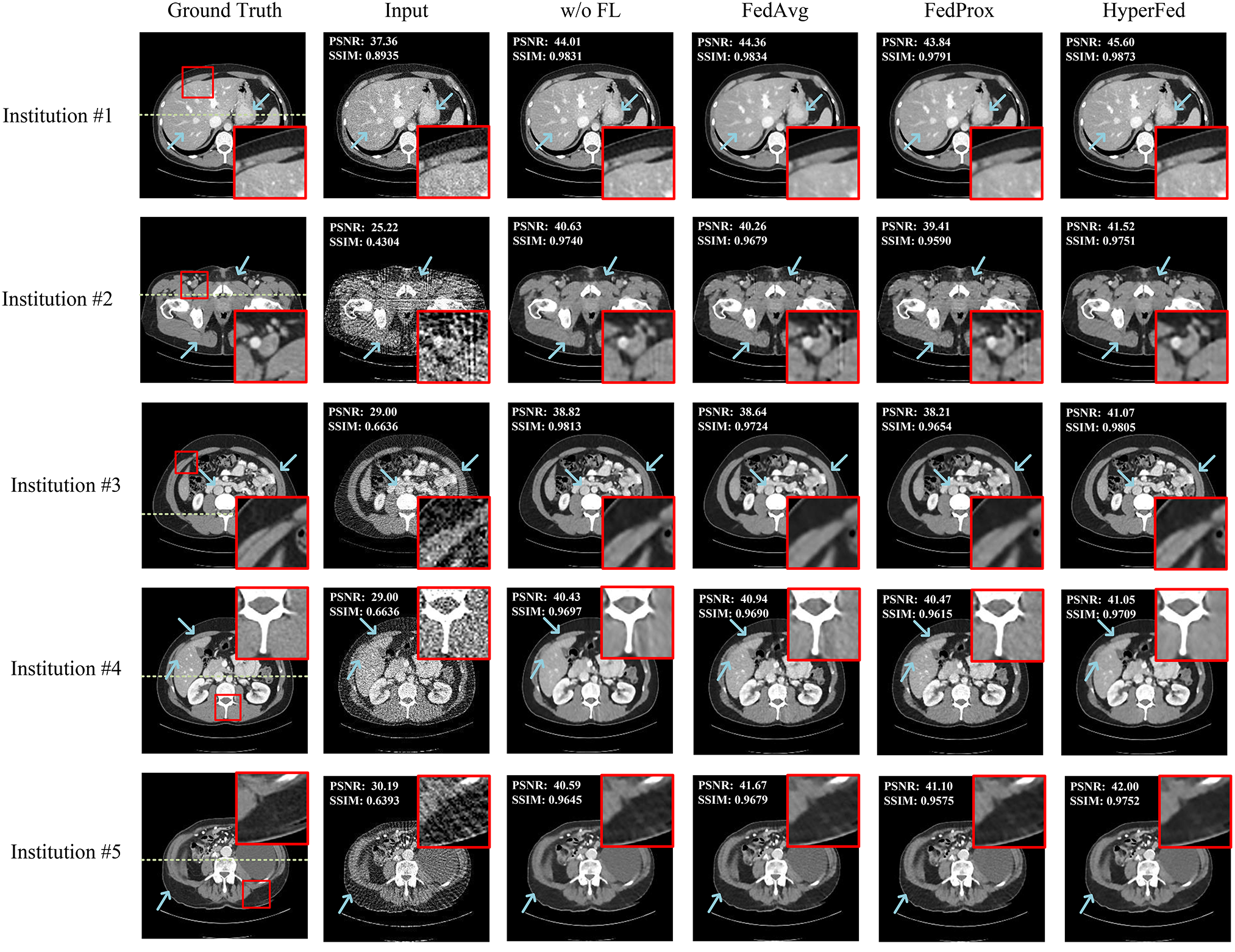}}
\caption{The results of w/o FL, FedAvg, FedProx and HyperFed under different geometries and dose levels for the reconstruction task. The display window is [-160, 240] HU.}
\label{fig9}
\end{figure*}

Fig. \ref{fig6} shows several typical slices denoised using different methods. It can be seen that the reconstructed images using FedAvg and FedProx still contain some noise or artifacts. The possible reason lies in that these methods are apt to extract global features rather than personalized features for different local models. Since the dataset for RED-CNN without FL for each institution only has 80 samples, which is relatively small, the details of the results are not well preserved. Benefiting from the Hypernetwork, HyperFed can generate more details aided by both the global features extracted by $F_{imag}$ and the local features extracted from $F_{hyper}$. Fig. \ref{fig7} shows the profiles along the green dotted lines in Fig. \ref{fig6} and it can be easily observed that the proposed method gains the closet result to the ground truth. Meanwhile, the proposed HyperFed achieves best scores in terms of both PSNR and SSIM in all the five cases. The architecture of HyperFed effectively splits the learning contents: the imaging network is responsible for learning some common imaging features from different institutions and the hypernetwork attempts to modulate the features of $F_{imag}$ to relieve the non-iid problem.

\begin{figure}[!t]
	\centerline{\includegraphics[width=\columnwidth]{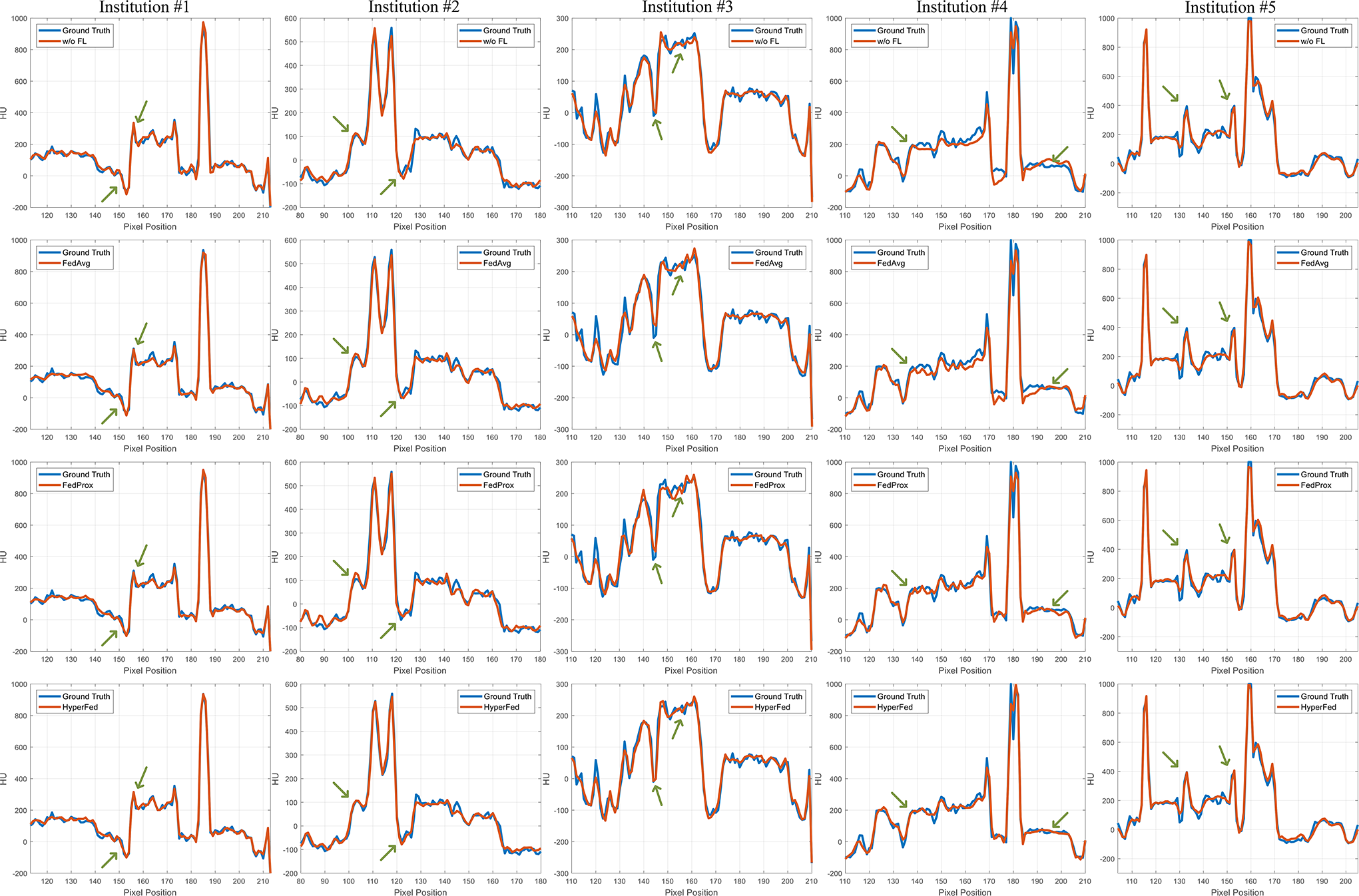}}
	\caption{Horizontal profiles of the results of w/o FL, FedAvg, FedProx and HyperFed.}
\label{fig10}
\end{figure}

\begin{figure}[!t]
	\centering
	\begin{minipage}[t]{0.32\columnwidth}
	\centering
	\includegraphics[width=\columnwidth]{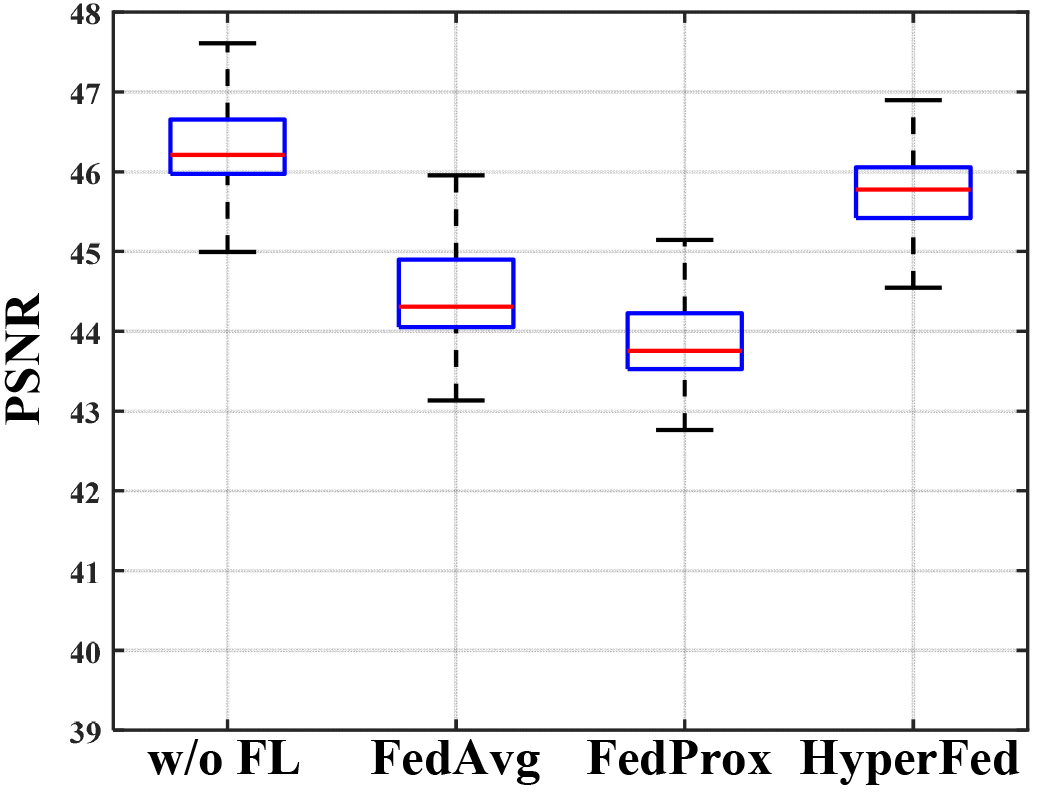}
	\centerline{(a)}
	\end{minipage}
	\begin{minipage}[t]{0.32\columnwidth}
	\includegraphics[width=\columnwidth]{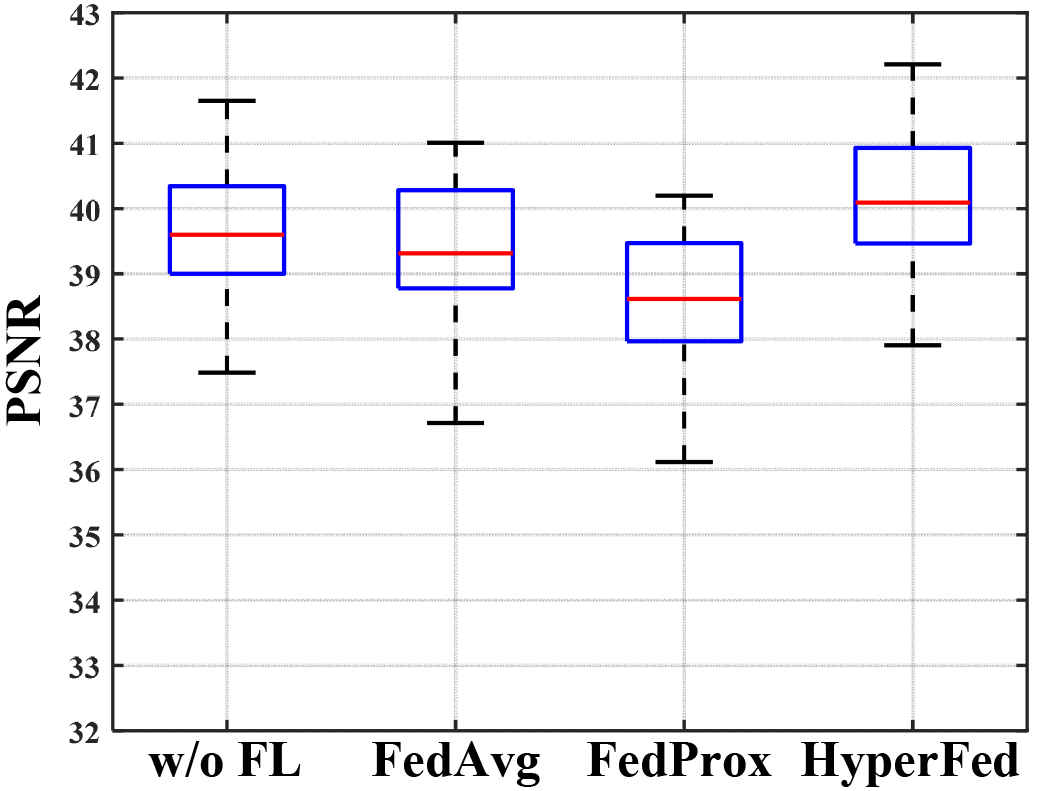}
	\centerline{(b)}
	\end{minipage}
	\begin{minipage}[t]{0.32\columnwidth}
	\includegraphics[width=\columnwidth]{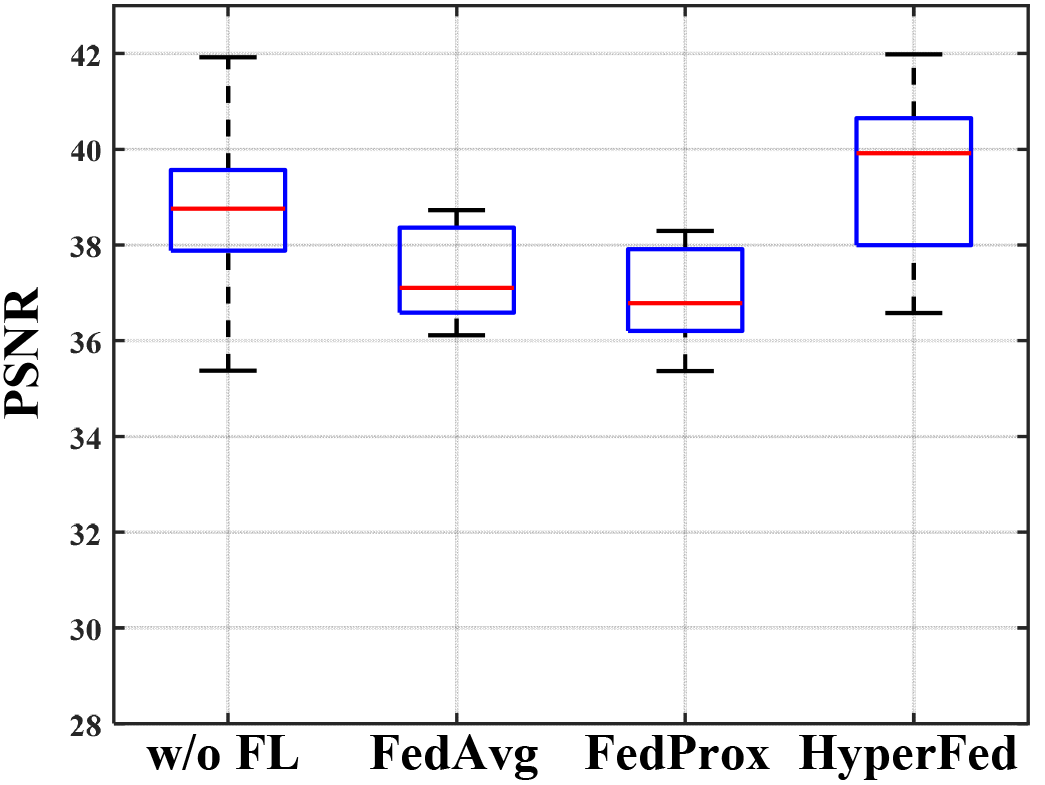}
	\centerline{(c)}
	\end{minipage}
\\
	\begin{minipage}[t]{0.32\columnwidth}
	\centering
	\includegraphics[width=\columnwidth]{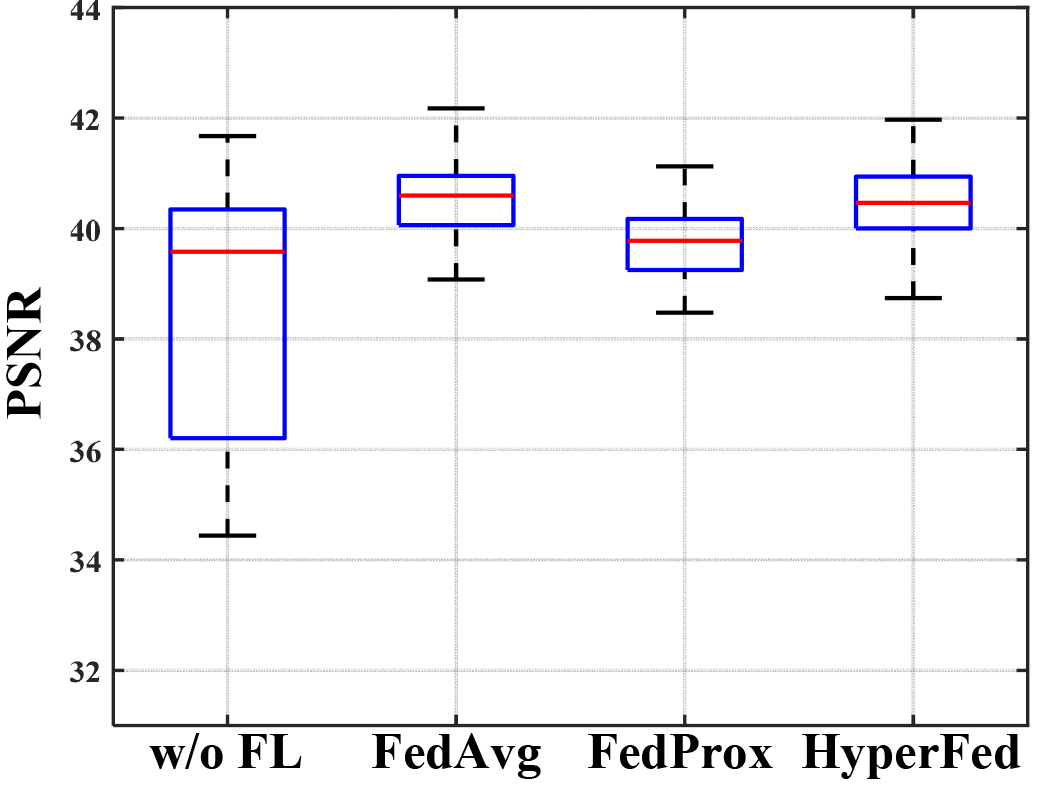}
	\centerline{(d)}
	\end{minipage}
	\begin{minipage}[t]{0.32\columnwidth}
	\includegraphics[width=\columnwidth]{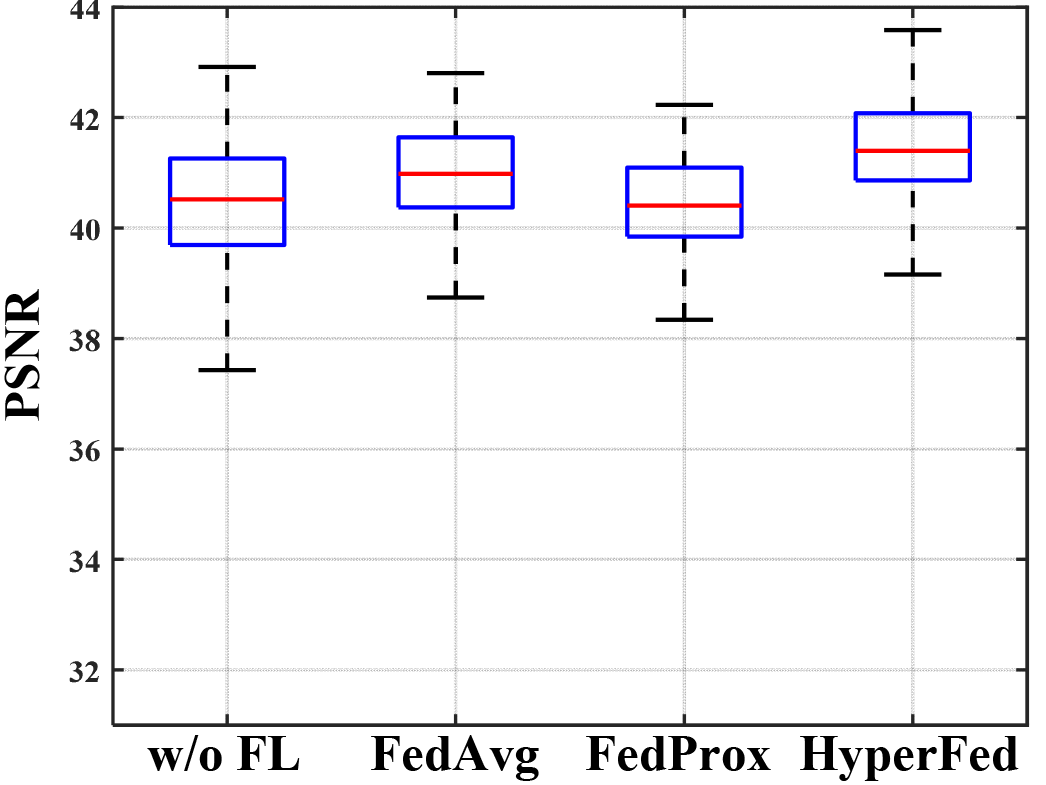}
	\centerline{(e)}
	\end{minipage}
	\begin{minipage}[t]{0.32\columnwidth}
	\includegraphics[width=\columnwidth]{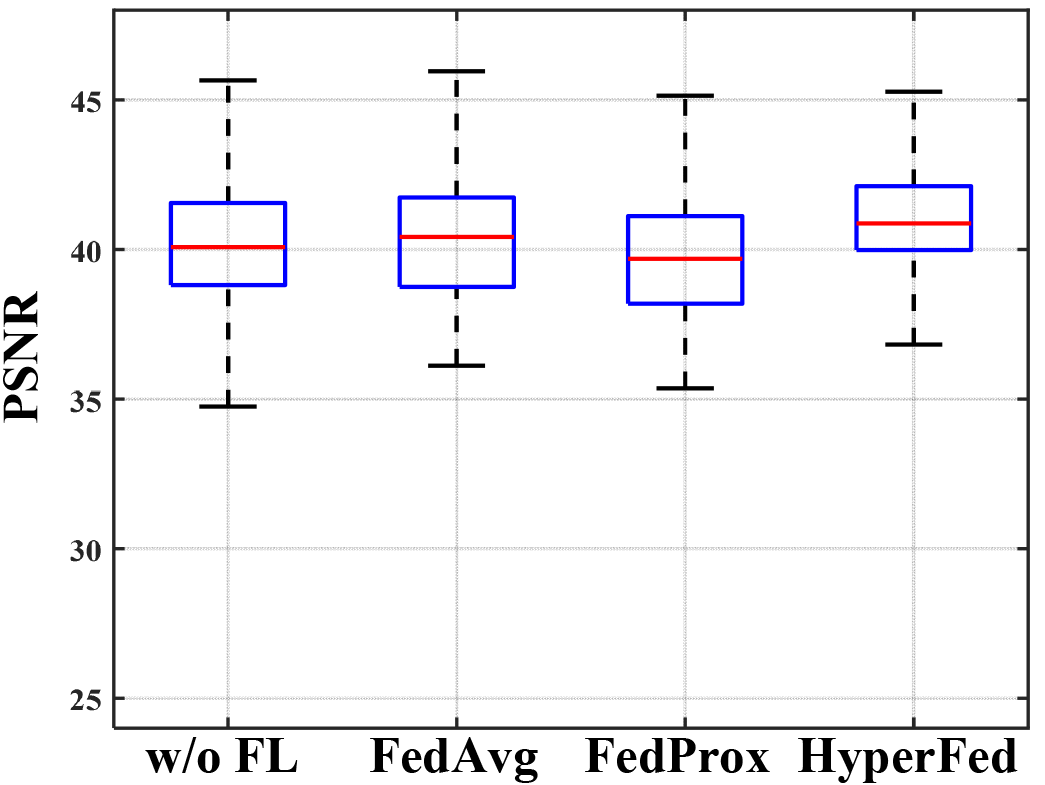}
	\centerline{(f)}
	\end{minipage}
\caption{The boxplots of PSNR based on w/o FL, FedAvg, FedProx and HyperFed for the reconstruction task. (a)-(g) represent the results of institution \#1 to \#5 respectively and (f) represents the average result of all institutions.}
\label{fig11}
\end{figure}


The quantitative results of the whole testing set are shown in Tab. \ref{tb3}. It can be seen that HyperFed achieved the best performance in most institutions compared with other methods. Besides, we offer boxplots in Fig. \ref{fig7} to demonstrate the stability of different methods. It can be noticed that HyperFed shows better stability and performance in comparison with other methods. The main reason is that the imaging network of HyperFed learns from different sources and has more data than the original RED-CNN, and the results support that our imaging network can extract stable features. Although other FL-based methods, both FedAvg and FedProx, are trained with mixed data, the performance is not satisfactory due to the diversity of data sources.

\subsection{Experiments for the Reconstruction Task}

In this subsection, LEARN, which belongs to unrolled reconstruction network, is included to validate the performance and generalization of the proposed HyperFed. Some typical slices reconstructed using different methods are shown in Fig. \ref{fig9} and Fig. \ref{fig10} demonstrates the profiles along the green dotted lines in Fig. \ref{fig9}. Similar to the results of post-processing task, it can be observed that there are still some noise and artifacts left in the results of FedAvg and FedProx, especially for the sparse-view cases. Due to the non-iid problem, simply combining with FL and adopting more training data from different sources can not improve the performance and even degrade the results a little. Our method still perform best for the unrolled reconstruction network in both visual inspection and quantitative matrices.

\begin{table*}[]
\centering
\caption{The Quantitative Results of PSNR and SSIM for the Reconstruction Task.}
\label{tb4}
\begin{tabular}{@{}lllllllll@{}}
\toprule
\centering
\multirow{2}{*}{} & \multicolumn{2}{c}{w/o   FL} & \multicolumn{2}{c}{FedAvg} & \multicolumn{2}{c}{FedProx} & \multicolumn{2}{c}{HyperFed} \\
                  & PSNR         & SSIM          & PSNR        & SSIM         & PSNR         & SSIM         & PSNR         & SSIM          \\ \midrule
Insitution \#1    & \textbf{46.26}                     & \textbf{0.9889}                   & 44.47                    & 0.9844                   & 43.87                    & 0.9771                   & 45.74                    & 0.9853                   \\
Insitution \#2    & 39.63                     & 0.9643                   & 39.35                    & 0.9604                   & 38.59                    & 0.9501                   & \textbf{40.17}                    & \textbf{0.9686}                   \\
Insitution \#3    & 38.79                     & \textbf{0.9800}                   & 37.37                    & 0.9698                   & 36.92                    & 0.9605                   & \textbf{39.47}                    & 0.9782                   \\
Insitution \#4    & 38.41                     & 0.9680                   & \textbf{40.51}                    & 0.9687                   & 39.73                    & 0.9578                   & 40.45                    & \textbf{0.9715}                   \\
Insitution \#5    & 40.46                     & 0.9714                   & 40.99                    & 0.9702                   & 40.42                    & 0.9618                   & \textbf{41.39}                    & \textbf{0.9750}                   \\
Overall           & 40.71                     & 0.9745                   & 40.54                    & 0.9707                   & 39.91                    & 0.9615                   & \textbf{41.44}                    & \textbf{0.9757}                   \\ \bottomrule
\end{tabular}
\end{table*}

\begin{figure*}[!t]
\centerline{\includegraphics[width=0.88\textwidth]{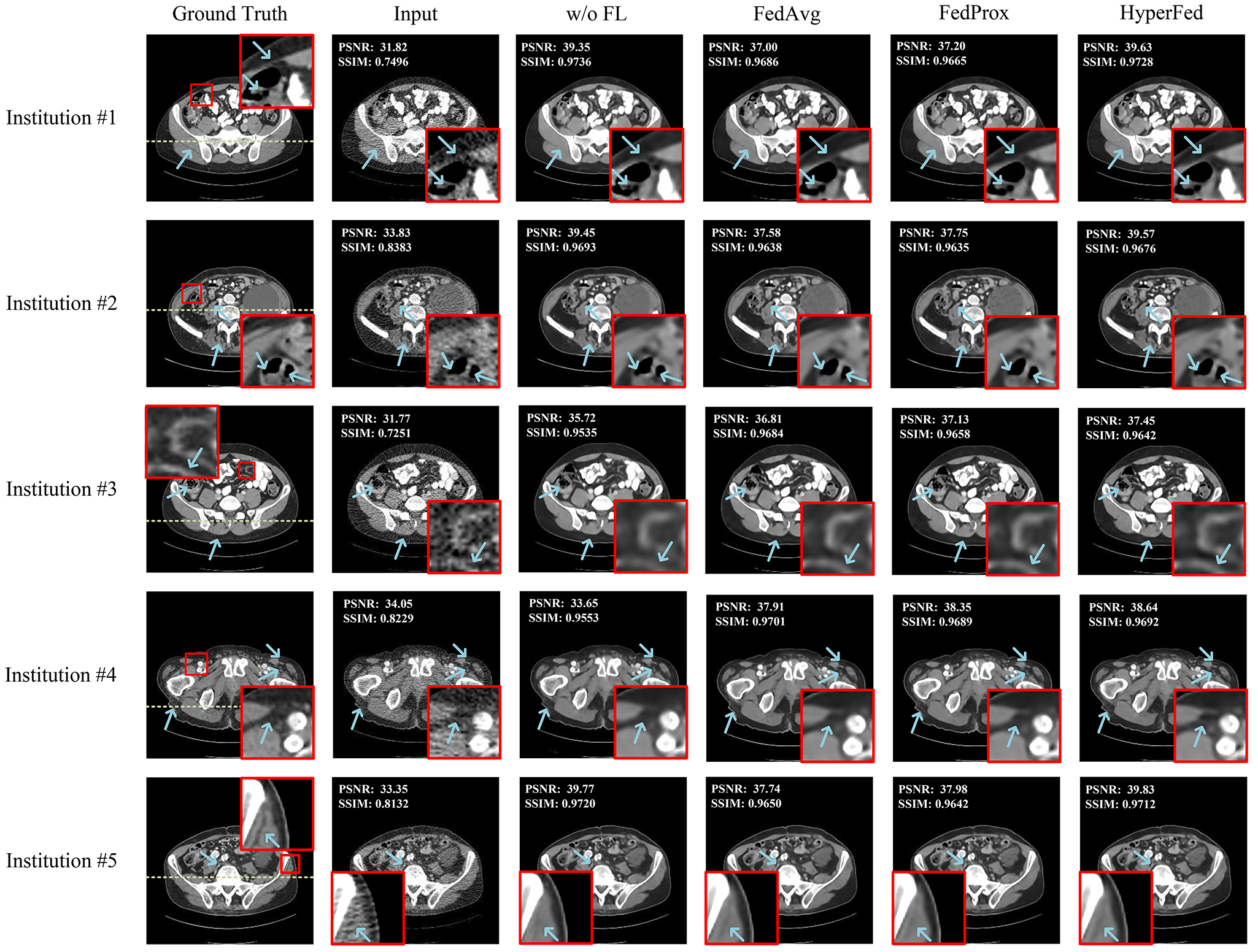}}
\caption{The results of w/o FL, FedAvg, FedProx and HyperFed under different geometries and dose levels trained by the large-scale training set. The display window is [-160, 240] HU.}
\label{fig12}
\end{figure*}
The quantitative results of the whole testing set are shown in Tab. \ref{tb4}. It can be seen that HpyerFed achieves the best performance in most cases compared with other methods. As shown in the boxplots in Fig. \ref{fig11}, HyperFed has the most stable performance in the comparison with other methods. The original LEARN without FL achieves good average quantitative performance and satisfactory imaging results in some institutions, but it cannot maintain its performance in each institution. Compared with the original LEARN, the performance of FL-based methods is more stable for different institutions. HyperFed integrates the advantages of CL and FL, achieving promising imaging performance in each institution.

From the above experiments for both post-processing and reconstruction tasks, we can see that it is difficult to achieve satisfactory imaging performance without considering non-iid problem and directly applying existing FL methods can not efficiently explore the power of big data. It is reasonable to divide the global optimization problem into the global imaging problem and the local feature domain adaption problem. Specifically, $F_{imag}$ is designed for learning some common imaging features from different institutions, and $F_{hyper}$ attempts to modulate the extracted features from $F_{imag}$ to alleviate the non-iid issue.

\subsection{Experiments on the Large-Scale Training Set}

\begin{table*}[]
\centering
\caption{The Quantitative Results of PSNR and SSIM for the Large-scale Training Set.}
\label{tb5}
\begin{tabular}{@{}lllllllll@{}}
\toprule
\centering
\multirow{2}{*}{} & \multicolumn{2}{c}{w/o   FL} & \multicolumn{2}{c}{FedAvg} & \multicolumn{2}{c}{FedProx} & \multicolumn{2}{c}{HyperFed} \\
                  & PSNR         & SSIM          & PSNR        & SSIM         & PSNR         & SSIM         & PSNR         & SSIM          \\ \midrule
Insitution \#1    & \textbf{39.30}            & 0.9385                   & 37.00                    & 0.9500                   & 37.33                    & 0.9484                   & 39.17                    & \textbf{0.9536}          \\
Insitution \#2    & \textbf{39.26}            & 0.9438                   & 36.89                    & 0.9501                   & 37.21                    & 0.9486                   & 39.07                    & \textbf{0.9539}          \\
Insitution \#3    & 35.59                     & 0.9424                   & 36.86                    & \textbf{0.9510}          & \textbf{37.20}           & 0.9491                   & 36.73                    & 0.9447                   \\
Insitution \#4    & 34.22                     & 0.9542                   & 37.15                    & \textbf{0.9551}          & 37.46                    & 0.9532                   & 37.85                    & 0.9524                   \\
Insitution \#5    & \textbf{40.17}            & 0.9458                   & 37.03                    & 0.9515                   & 37.31                    & 0.9495                   & 39.41                    & \textbf{0.9555}          \\
Overall           & 37.71                     & 0.9450                   & 36.99                    & 0.9515                   & 37.30                    & 0.9498                   & \textbf{38.44}           & \textbf{0.9520}          \\ \bottomrule
\end{tabular}
\end{table*}
As shown in the previous experiments, some FL-based methods and the original imaging models without FL cannot work well with small local training set. In this subsection, we compare the performance of these methods trained with a large-scale local data set. RED-CNN is used as the backbone network. The other settings about hyperparameters are exactly the same as the previous experiments and the only difference is the size of training set. There are total 1000 images are randomly selected as the training set and these images are randomly divided into five groups on average. For each institution, 200 images are chosen to simulate the training set in each local image domain. As shown in Figs. \ref{fig12} and \ref{fig10}, FL-based methods benefit more from the increase of dataset size, and their performances improve significantly. It indicates that the non-iid problem can be relieved by adding more training samples. Similar to the above experiments, the original RED-CNN cannot recover all the details since it does not have any features from other data domains. Compared with other methods, HyperFed can still keep its remarkable performance, which supports that the proposed architecture can improve the imaging quality regardless of the size of training set.

\begin{figure}[!t]
	\centerline{\includegraphics[width=\columnwidth]{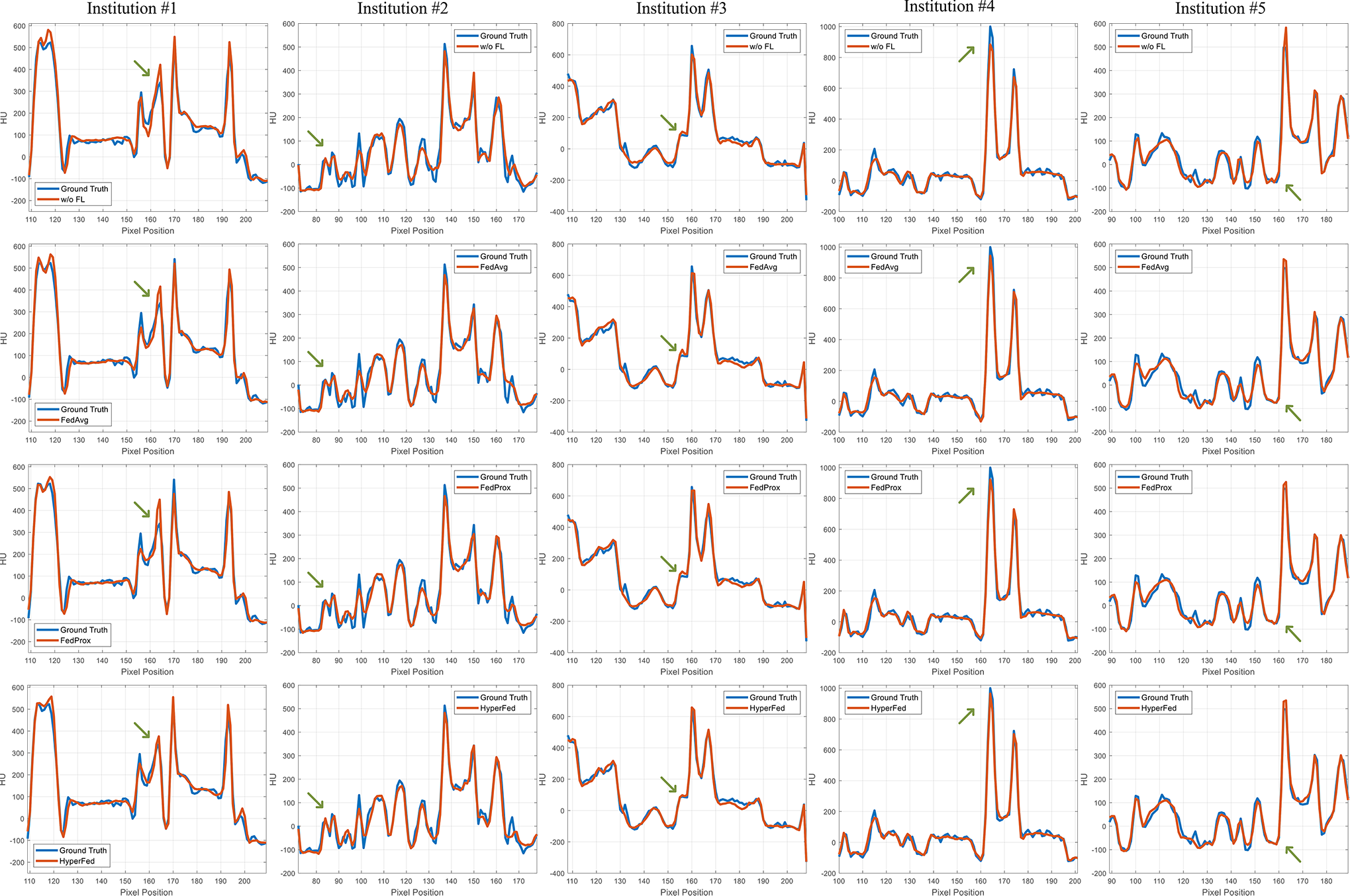}}
	\caption{Horizontal profiles of the results of w/o FL, FedAvg, FedProx and HyperFed.}
\label{fig13}
\end{figure}

\begin{figure}[!t]
	\centering
	\begin{minipage}[t]{0.32\columnwidth}
	\centering
	\includegraphics[width=\columnwidth]{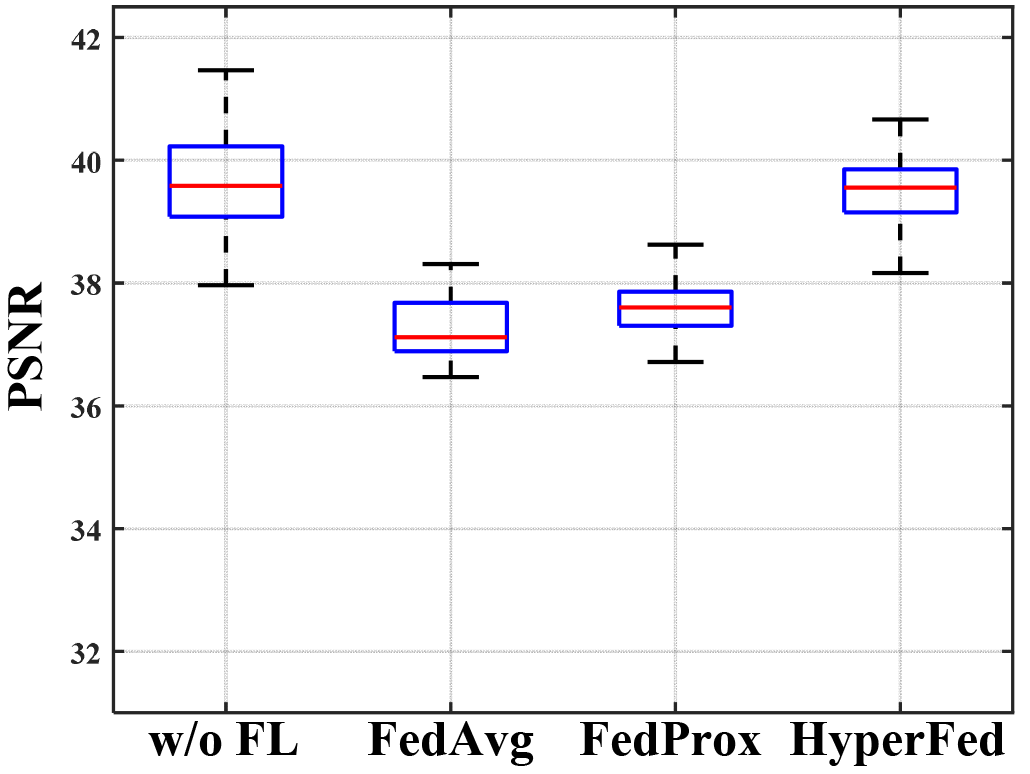}
	\centerline{(a)}
	\end{minipage}
	\begin{minipage}[t]{0.32\columnwidth}
	\includegraphics[width=\columnwidth]{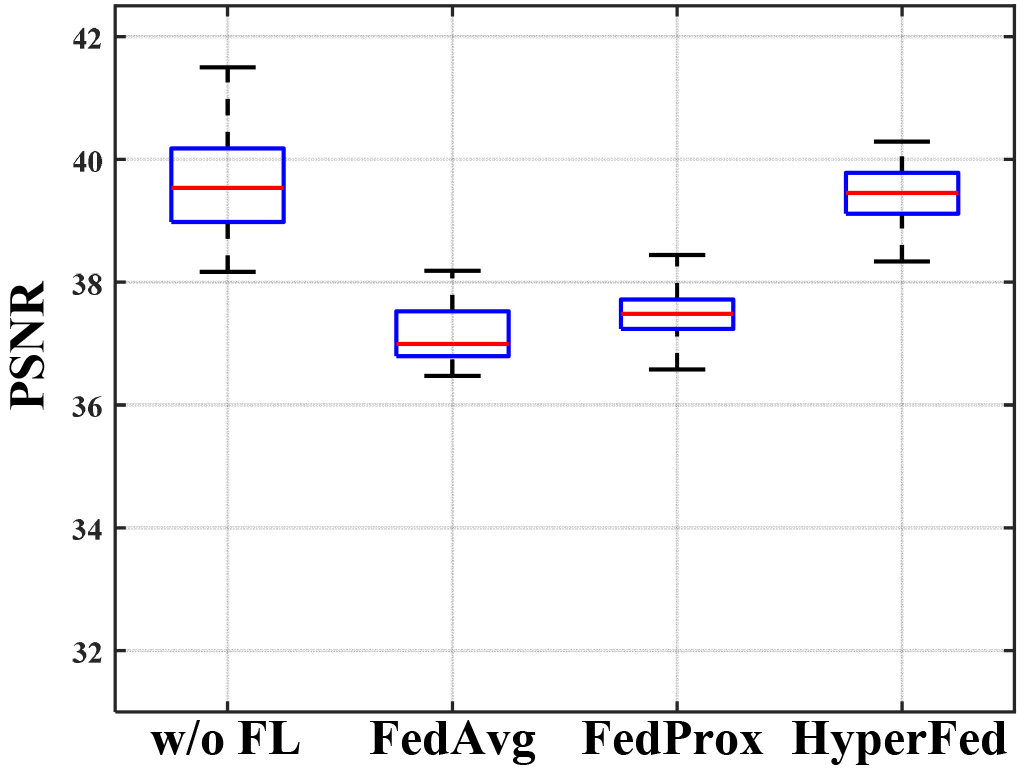}
	\centerline{(b)}
	\end{minipage}
	\begin{minipage}[t]{0.32\columnwidth}
	\includegraphics[width=\columnwidth]{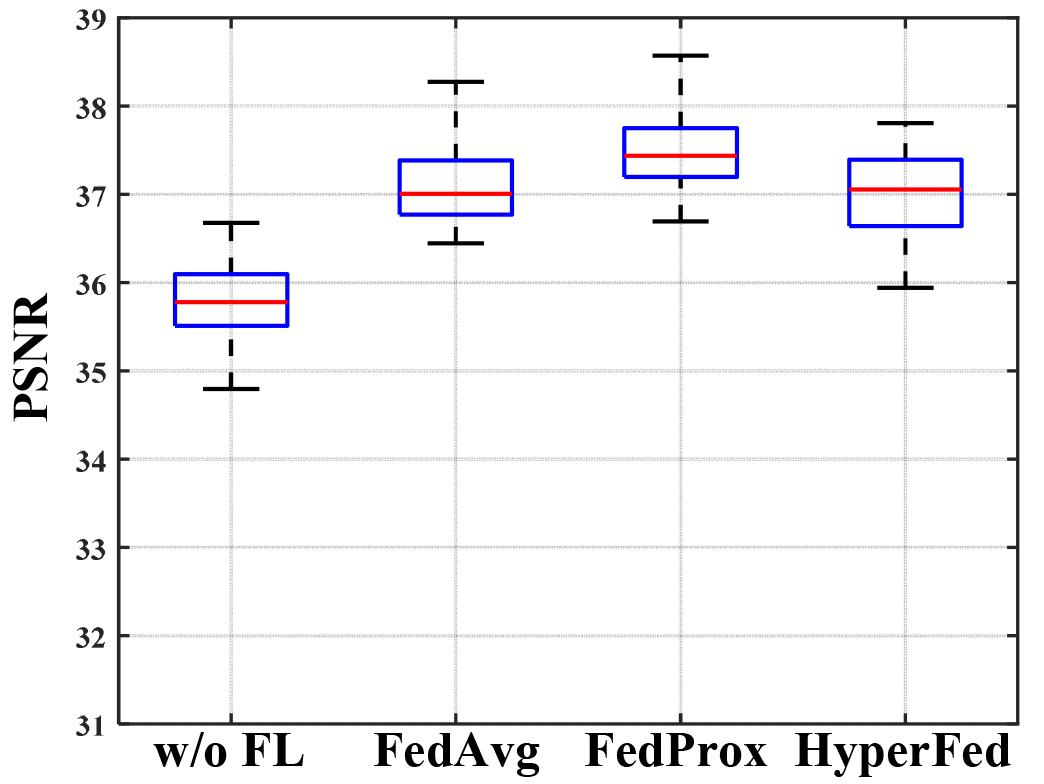}
	\centerline{(c)}
	\end{minipage}
\\
	\begin{minipage}[t]{0.32\columnwidth}
	\centering
	\includegraphics[width=\columnwidth]{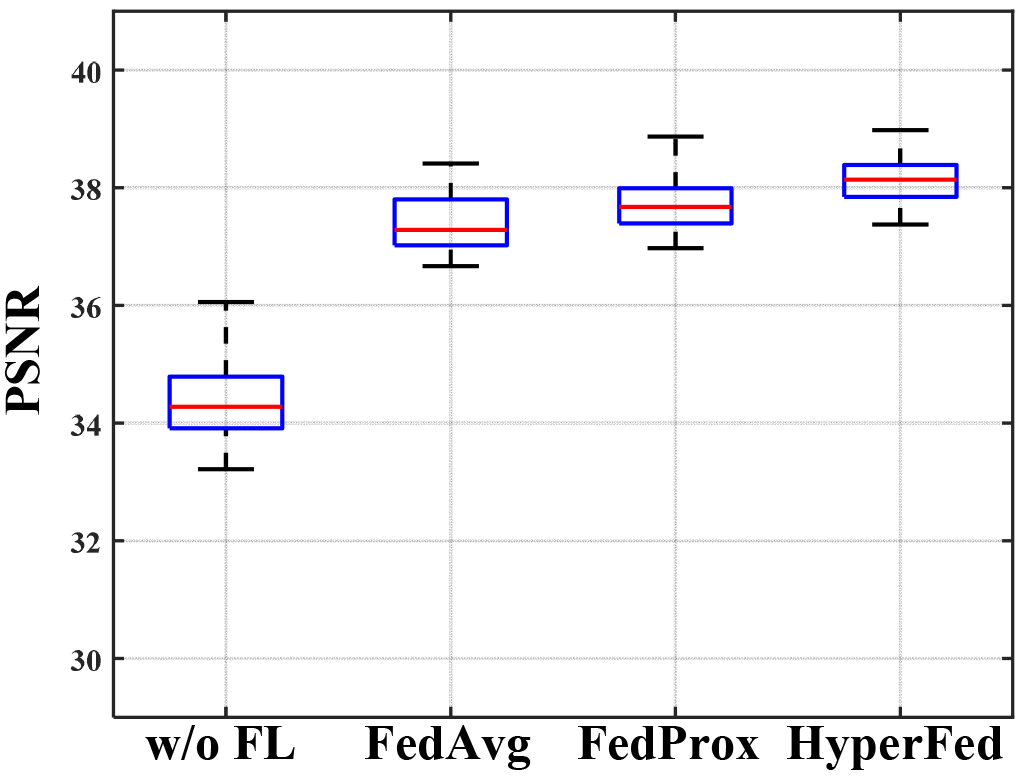}
	\centerline{(d)}
	\end{minipage}
	\begin{minipage}[t]{0.32\columnwidth}
	\includegraphics[width=\columnwidth]{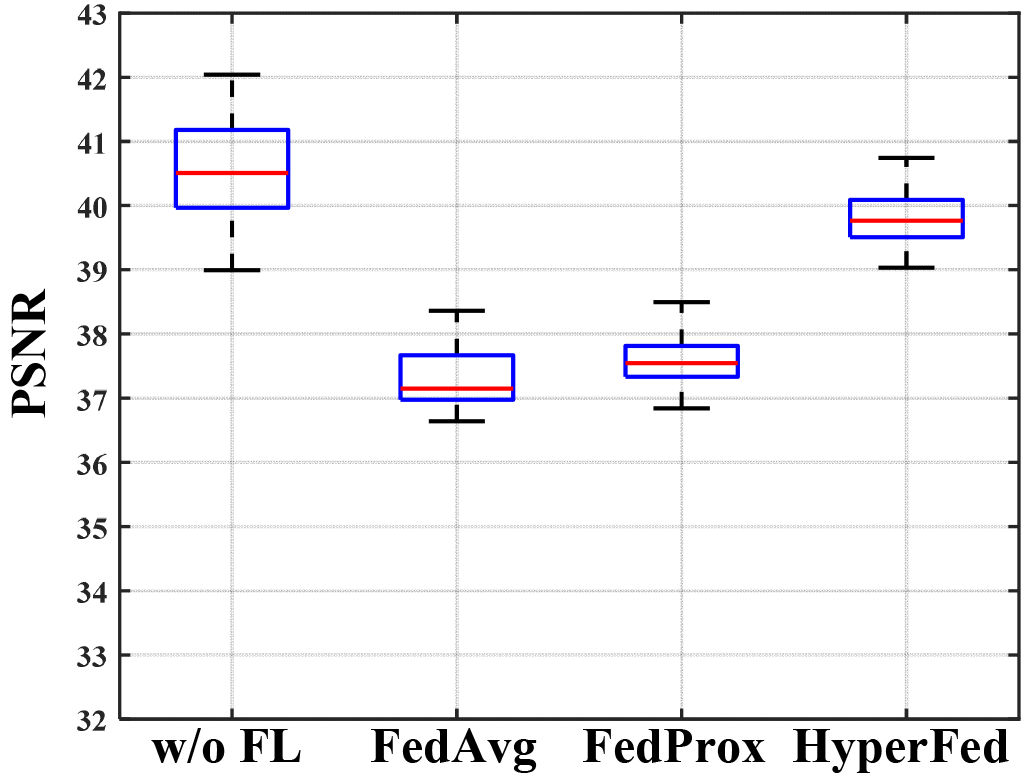}
	\centerline{(e)}
	\end{minipage}
	\begin{minipage}[t]{0.32\columnwidth}
	\includegraphics[width=\columnwidth]{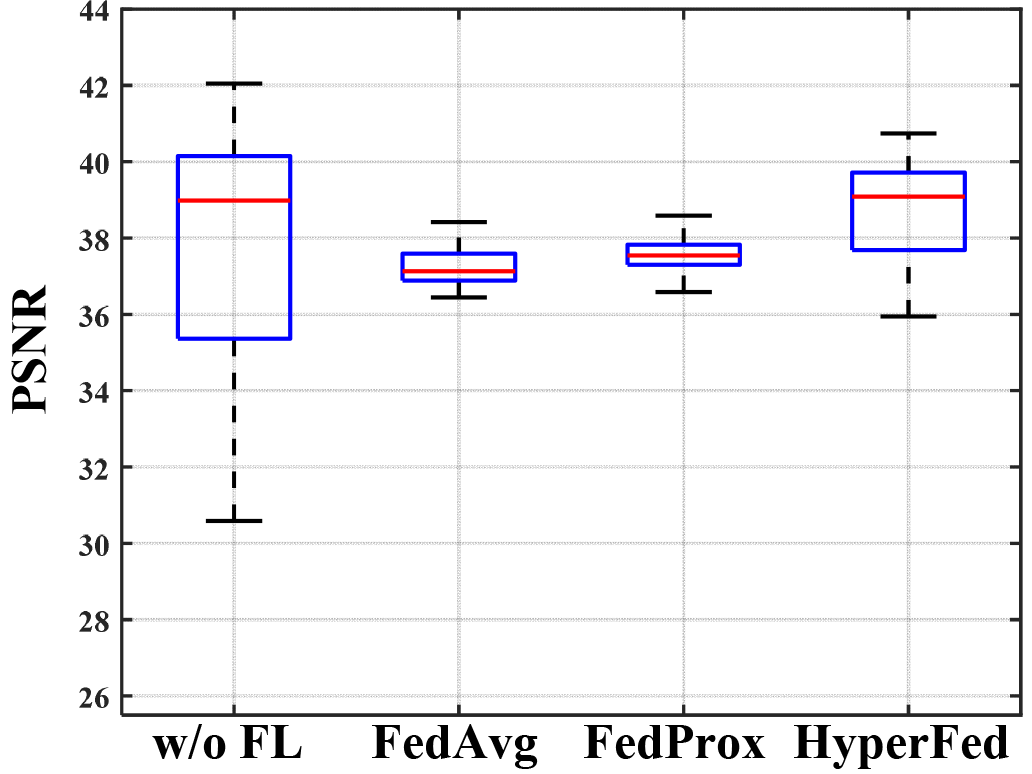}
	\centerline{(f)}
	\end{minipage}
\caption{The boxplots of PSNR based on w/o FL, FedAvg, FedProx and HyperFed on the large-scale training set. (a)-(g) represent the results of institution \#1 to \#5 respectively and (f) represents the average result of all institutions.}
\label{fig14}
\end{figure}

Tab. \ref{tb5} shows the quantitative results, and all the methods can achieve improved performance due to the increase of training samples. In this case, HyperFed still has competitive performance compared with other methods. Although the original RED-CNN achieves better results in some cases, its stability is not improved as shown in the profiles and boxplots in Figs. \ref{fig13} and \ref{fig14}, respectively. FL-based methods maintain their stability in all the experiments, but the problem for these methods is that they cannot adjust extracted features based on the characteristics of different domains. In other words,  the cost of stability is the personalization. Classical FL methods cannot work well for some specific cases and lack of personalization leads to unsatisfactory reconstruction. The reason lies in that these methods attempt to minimize the average imaging loss of different institutions rather than the best local imaging reconstruction. All experiments have shown that the proposed HyperFed can balance well between the stability and the imaging performance in the learning process. HyperFed integrates the advantages of both training modes and achieves stable and competitive performance.

\section{Conclusion}

Current CT imaging networks are CL-based without considering any privacy issue. Due to the inevitable non-iid problem among different institutions in CT imaging, classical FL methods cannot achieve satisfactory performance. To relieve the non-iid problem, we propose a novel hypernetwork-based personalized federated learning framework for multi-institution CT reconstruction, dubbed as HyperFed. The basic architecture of HyperFed is composed of an institution-specific hypernetwork and a global-sharing imaging network. The global sharing imaging network ensures the reconstruction is stable and the hypernetwork modulates the feature domains to achieve better performance. Experimental results show that the method w/o FL can personalize CT reconstruction, but some details are lost when the number of training samples is small. FL-based methods can keep stable performance and recover more details for all institutions, but they cannot reconstruct high-quality CT images for each institution. The proposed HyperFed effectively balances the tradeoff between both factors and it can reconstruct high-quality CT images for each institution compared with other methods in both qualitative and quantitative aspects.


\bibliographystyle{ieeetr}
\bibliography{ref}

\end{document}